\documentclass[fleqn,usenatbib]{mnras}

\usepackage[T1]{fontenc}

\DeclareRobustCommand{\VAN}[3]{#2}
\let\VANthebibliography\thebibliography
\def\thebibliography{\DeclareRobustCommand{\VAN}[3]{##3}\VANthebibliography}

\usepackage{graphicx}	
\usepackage{amsmath}	
\usepackage{amssymb}	

\title[Magnetic activities on IM~Peg and $\sigma$~Gem]{Magnetic activities on two single-lined RS~Canum Venaticorum binaries IM~Pegasi and $\sigma$~Geminorum}
\author[D. Cao et al.]
            {Dongtao Cao,$^{1, 2}$\thanks{E-mail: dtcao@ynao.ac.cn, shenghonggu@ynao.ac.cn}
             Shenghong Gu,$^{1, 2, 3}$\footnotemark[1]
             F. Grundahl,$^{4}$
             and P. L. Pall\'{e},$^{5, 6}$ 
\\
$^{1}$Yunnan Observatories, Chinese Academy of Sciences, Kunming 650216, China\\
$^{2}$Key Laboratory for the Structure and Evolution of Celestial Objects, Chinese Academy of Sciences, Kunming 650216, China\\
$^{3}$School of Astronomy and Space Science, University of Chinese Academy of Sciences, Beijing 101408, China\\
$^{4}$Stellar Astrophysics Centre, Department of Physics and Astronomy, Aarhus University, NyMunkegade 120, 8000 Aarhus C, Denmark\\
$^{5}$Instituto de Astrof\'{i}sica de Canarias, 38205 La Laguna, Tenerife, Spain\\
$^{6}$Universidad de La Laguna, Departamento de Astrof\'{i}sica, 38206 La Laguna, Tenerife, Spain
}

\date{Accepted XXX. Received YYY; in original form ZZZ}

\pubyear{2015}

\begin{document}
\label{firstpage}
\pagerange{\pageref{firstpage}--\pageref{lastpage}}
\maketitle

\begin{abstract}
We present the study on continuous high-resolution spectroscopic observations of two long-period single-lined RS Canum Venaticorum (RS CVn) binary stars IM~Pegasi (IM~Peg) and $\sigma$~Geminorum ($\sigma$~Gem), obtained with the Hertzsprung SONG telescope during the 2015--2016 season. Chromospheric activity indicators H$_{\alpha}$, $\mbox{Na~{\sc i}}$ D$_{1}$, D$_{2}$ doublet, $\mbox{He~{\sc i}}$ D$_{3}$, and H$_{\beta}$ lines have been analyzed by using the spectral subtraction technique. The expected chromospheric emission features in the H$_{\alpha}$, $\mbox{Na~{\sc i}}$ D$_{1}$, D$_{2}$ doublet, and H$_{\beta}$ lines confirm that both of two stars are very active systems. In the spectra, the $\mbox{He~{\sc i}}$ D$_{3}$ line had been always detected in absorption feature. Although the behavior of chromospheric activity indicators is very similar for both stars, the activity level of IM~Peg is much stronger than that of $\sigma$~Gem. Moreover, the EW variations of the H$_{\alpha}$, $\mbox{He~{\sc i}}$ D$_{3}$, and H$_{\beta}$ line subtractions correlate well and show different behavior among different orbital cycles, which indicates the presence and evolution of activity longitudes over the surface of two stars. Furthermore, the subtracted H$_{\alpha}$ line profile is usually asymmetric. The red-shifted excess absorption features could be interpreted as a strong down-flow of cool absorbing material, while the blue-shifted emission component is probably caused by up-flow of hot materials through microflare events.
\end{abstract}

\begin{keywords}
                   stars: magnetic field ---
                   stars: chromospheres ---
                   stars: activity ---
                   stars: individual: IM~Peg ---
                   stars: individual: $\sigma$~Gem
\end{keywords}

\section{Introduction}
\indent
A wide range of solar-type activity phenomena, including starspots, plages, flares, prominences, etc, have been widely observed in many cool stars \citep{schrijver2000}. It is commonly accepted that all of these active phenomena arise from a powerful magnetic dynamo generated by the interplay between the turbulent motion in the convection zone and the stellar differential rotation, in a manner similar to the solar case. Thus, the solar activity paradigm provides a better reference to investigate magnetic activity phenomena encountered in cool stars. At Yunnan Observatories, we have begun a long-term high-resolution spectroscopic monitoring project on a number of RS~CVn-type binary systems, to study their magnetic activity (detecting optical flares, searching for prominence-like events, exploring the rotational modulation of chromospheric activity, and investigating the evolution of active regions) based on the information derived through several optical chromospheric activity indicators \citep{cao2012, cao2015, cao2017, cao2020}.\\
\indent
IM~Peg (=~HD~216489~=~HR~8703) is a long-period single-lined spectroscopic binary system, which was classified as a RS CVn-type star \citep{hall1976}. The system consists of a K2~III primary star and an unseen secondary companion in an almost circular orbit with a period of about 24.6 days \citep{berdyugina1999}. The K2~III primary star shows significant starspots displayed by photometric light curves (e.g., \citealt{Strassmeier1997}) and surface maps with Doppler imaging technique (e.g., \citealt{berdyugina2000}). Moreover, the K2~III primary has intense chromospheric emission and rotational modulation of the $\mbox{Ca~{\sc ii}}$ IRT, H$_{\alpha}$, and $\mbox{Mg~{\sc ii}}$ h\&k lines \citep{Huenemoerder1990, Dempsey1993, Dempsey1996, Frasca1994, Biazzo2006}. \citet{Huenemoerder1990} found that the H$_{\alpha}$ intensity was possibly correlated with the orbital phase. In addition, \citet{Biazzo2006} derived a tight anti-correlation between the chromospheric H$_{\alpha}$ emission and the photospheric temperature modulations, which indicates a close spatial association between photospheric starspots and chromospheric plage regions. Furthermore, it is noteworthy that there were some interesting excess absorption features in the H$_{\alpha}$ line profile at some phases in their observations.\\
\indent
$\sigma$~Gem (= 75~Gem = HD~62044 = HR~2973) is also classified as a bright single-lined spectroscopic binary star with a relatively long orbital period of about 19.6 days, consisting of a K1~III primary component and an unseen secondary companion \citep{Ayres1984, Duemmler1997}. \citet{Roettenbacher2015} detected the secondary companion with a primary-to-secondary $H$-band flux ratio of 270$\pm$70 via interferometric observations.  As a member of RS CVn-type stars, $\sigma$~Gem shows starspots detected by photometric observations (see, e.g., \citealt{Hall1977, Zhang1999}) and mapped with Doppler imaging technique (e.g., \citealt{Hatzes1993, Kovari2001, Roettenbacher2017, Korhonen2021}). \citet{Roettenbacher2017} also derived interferometric image of starspots on $\sigma$~Gem and investigated the first comparisons of the interferometric, spectroscopic, and photometric imaging techniques. Moreover, chromospheric emission has been found in the H$_{\alpha}$, H$_{\beta}$, $\mbox{Ca~{\sc ii}}$ H and K, and $\mbox{Ca~{\sc ii}}$ IRT lines \citep{Strassmeier1990, Frasca1994, Zhang1999, montes2000}, revealing the unusual manifestations of $\sigma$~Gem with high chromospheric activity. Additionally, \citet{Zhang1999} found that the chromospheric H$_{\alpha}$ emission vary with orbital phase and shows obvious correlation with the photospheric spot regions, which implies the spatial association between activity regions at different atmospheric levels.\\
\indent
In this paper, we present the investigations on IM~Peg and $\sigma$~Gem, based on continuous high-resolution spectroscopic observations obtained from 2015 to 2016. The details of our spectroscopic observations and data reduction are given in the Sec.~2, and the procedure of the spectral analysis is described in Sec.~3. In Sec.~4, the behavior of chromospheric activity indicators and the variation of chromospheric activity of IM~Peg and $\sigma$~Gem are respectively discussed. In addition, we make a discussion to the asymmetric H$_{\alpha}$ feature. Finally, we give a summary and state the conclusions of our study in Sec.~5.\\
\section{Spectroscopic observations and data reduction}
\begin{figure*}
    \centering
     \includegraphics[width=17.5cm,height=8.75cm, angle=270]{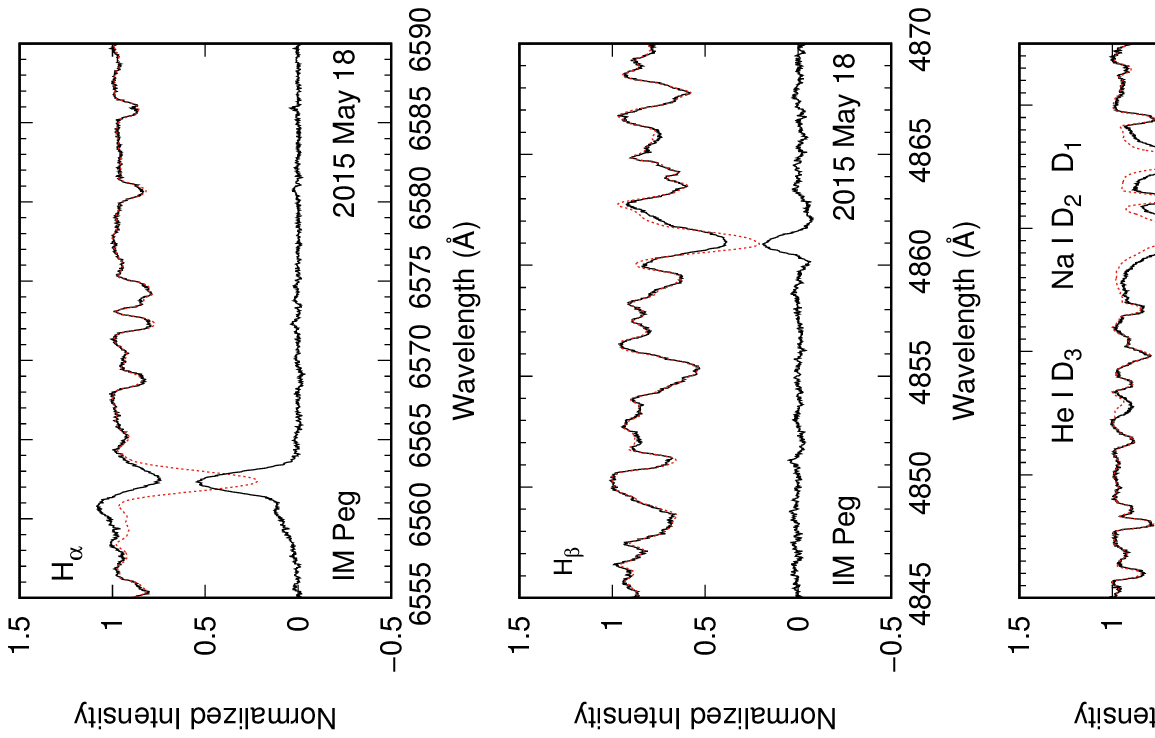}
     \includegraphics[width=17.5cm,height=8.75cm, angle=270]{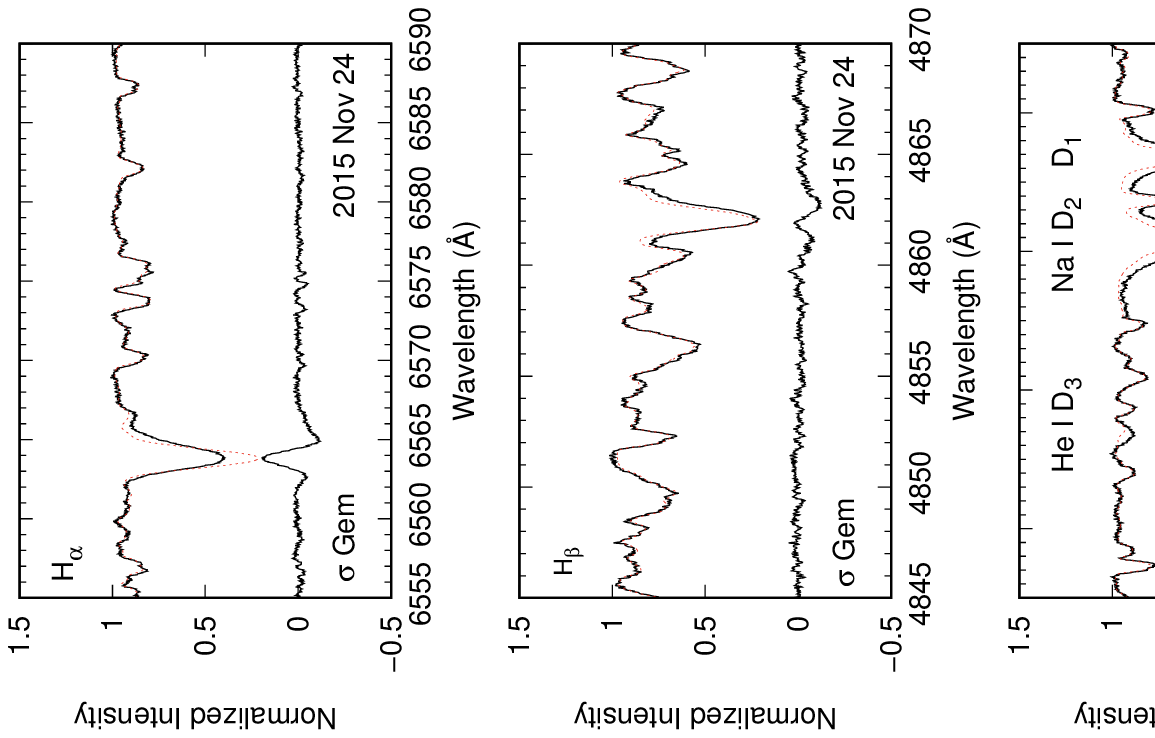}
     \caption{Examples of the observed, synthesized, and subtracted spectra for the chromospheric activity indicators H$_{\alpha}$, $\mbox{Na~{\sc i}}$ D$_{1}$, D$_{2}$ doublet, 
                        $\mbox{He~{\sc i}}$ D$_{3}$, and H$_{\beta}$ line spectral regions of IM~Peg (left panels) and $\sigma$~Gem (right panels). For 
                        each panel, the upper solid-line is the observed spectrum, the dotted line represents the synthesized spectrum and the lower spectrum 
                        is the subtracted one.}
      \label{fig1}
\end{figure*}

\indent
During the 2015--2016 observing season, high-resolution spectroscopic observations of IM~Peg and $\sigma$~Gem were carried out with the prototype node of the Stellar Observations Network Group (SONG, \citealt{Grundahl2006, Grundahl2009, Andersen2014, Andersen2019}), SONG-OT \citep{Uytterhoeven2012}, at Teide Observatory in Tenerife, Spain. The SONG project aims at constructing a global network of 1~meter robotic telescopes, equipped with high-resolution echelle spectrographs in the coud\'{e} path of the telescopes \citep{Grundahl2017}. The echelle spectrograph has resolving power ($R$~=~$\lambda$/$\Delta\lambda$) from 35000 to 110000 with different slit width and covers the wavelength range from 4400 to 6900~\AA. Moreover, a $2048 \times 2048$ pixels CCD detector was used to record spectra. For our observations, we chose slit~\#5 with a resolving power ($\lambda$/$\Delta\lambda$) of about 77000.\\ 
\indent
The observing log of IM~Peg is listed in Table~\ref{tab1}, which includes the date, the heliocentric Julian date (HJD), and orbital phase calculated with the ephemeris:
\begin{equation}
$HJD$ = 2,450,342.905 + 24^{d}.64877 \times E,
\end{equation}
from \citet{Marsden2005}, where the epoch corresponds to the conjunction with the K2~III primary star in the back of the binary system. In total, 63 spectra of IM~Peg were obtained with exposure time of 600 seconds during the observations. Moreover, the detailed log of $\sigma$~Gem could be found in Table~\ref{tab2}, in which the orbital phases are calculated with the ephemeris:
\begin{equation}
$HJD$ = 2,450,388.36853 + 19^{d}.60447 \times E,
\end{equation}
from \citet{Kovari2001}, where the epoch corresponds to maximum positive velocity of the K1~III primary star. For the star $\sigma$~Gem, we obtained 271 spectra during the observations and each individual spectrum had an exposure time of 180 seconds.\\ 
\indent
In addition, besides our two main target stars, some rapidly rotating early-type stars and slowly rotating inactive stars with the same spectral types and luminosity classes as the components of IM~Peg and $\sigma$~Gem were also observed with the same instrumental setup during the observations. The spectra of early-type stars were used as telluric absorption line templates whereas the inactive stars were used as references in the synthesized spectral subtraction of chromospheric activity indicators. The observations of early-type stars were approximately repeated once a month during our continuous observations. \\
\indent
The 1D spectra of our observations were fully automatically extracted with the SONG pipeline \citep{Corsaro2012, Antoci2013}. Then, we normalized them by using low-order polynomial fit to the observed continuum with the CONTINNUM task in the IRAF$^{1}$\footnotetext[1]{IRAF is distributed by the National Optical Astronomy Observatories, which is operated by the Association of Universities for Research in Astronomy (AURA), Inc., under cooperative agreement with the National Science Foundation.} package. If the continuum level is inconsistent between the different epochs, we can refit them and therefore made the continuum as consistent as possible. Moreover, for most of our observations during which the H$_{2}$O telluric absorption lines were very heavy in the chromospheric activity line regions of interest, two rapidly rotating early-type stars HR~7894 (B5~IV, $vsini$ = 330 km~s$^{-1}$) and HR~1051 (B8~V, $vsini$ = 295 km~s$^{-1}$) were used as telluric line templates for IM~Peg and $\sigma$ Gem, respectively. The telluric lines in the target spectra were eliminated by means of the templates with an interactive procedure in the IRAF package, as described in detail by \citet{gu2002}.\\
\indent
Examples of the normalized H$_{\alpha}$, $\mbox{Na~{\sc i}}$ D$_{1}$, D$_{2}$ doublet, $\mbox{He~{\sc i}}$ D$_{3}$, and H$_{\beta}$ line profiles of IM~Peg and $\sigma$~Gem obtained during the observations can be found in Fig.~\ref{fig1}, respectively, where the star name and observing time are also marked.\\
\begin{figure*}
    \centering
     \includegraphics[width=17.5cm,height=8.75cm, angle=270]{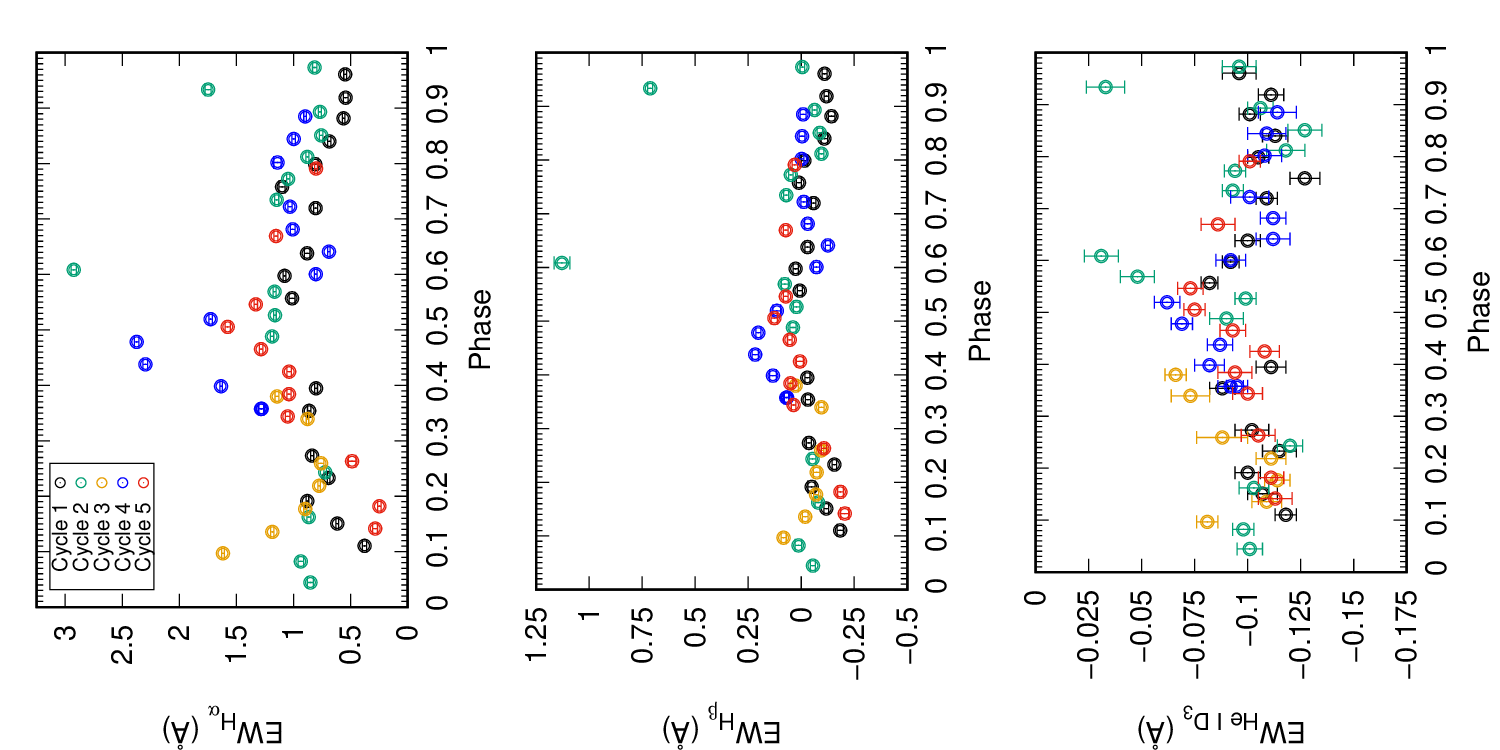}
     \includegraphics[width=17.5cm,height=8.75cm, angle=270]{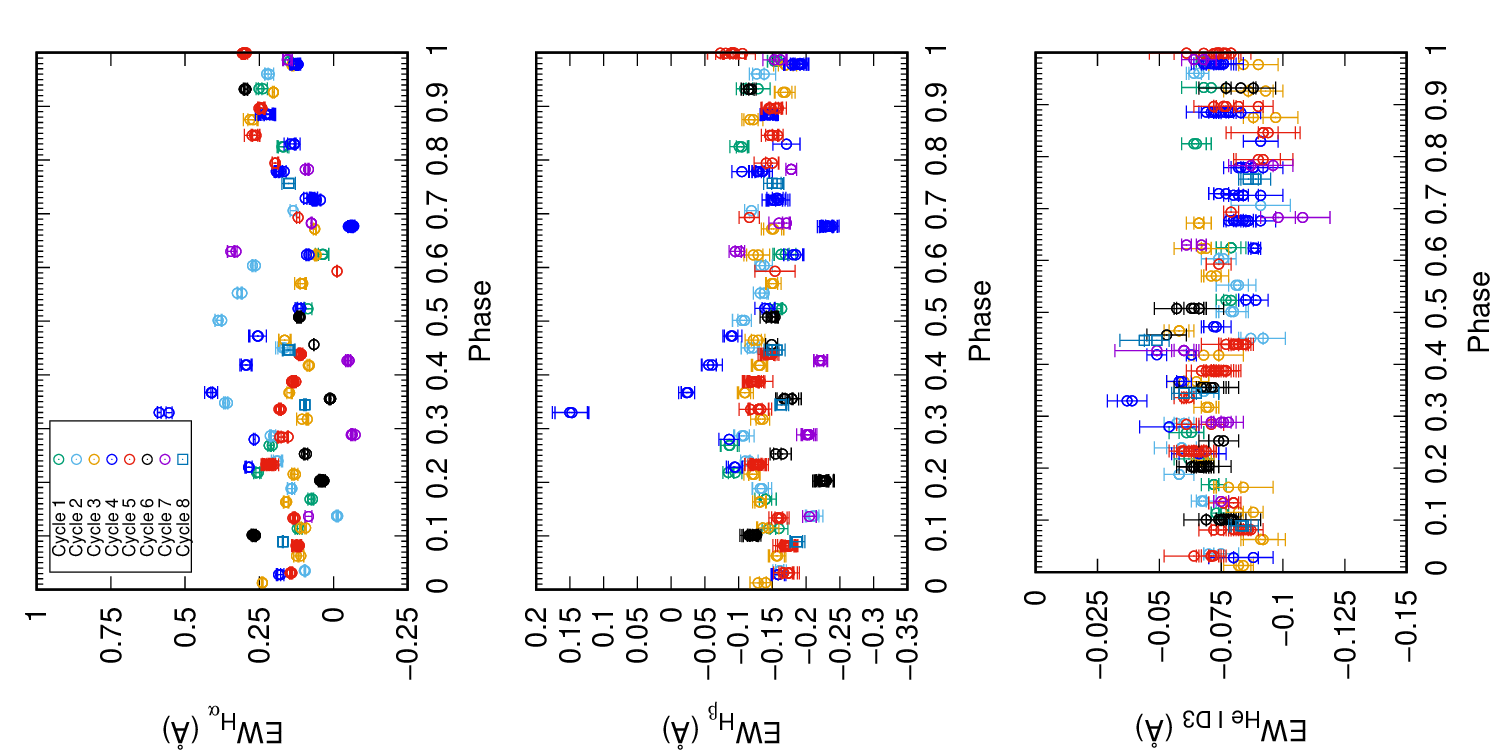}
     \caption{EWs of the subtraction versus orbital phase for the chromospheric activity indicators H$_{\alpha}$, $\mbox{He~{\sc i}}$ D$_{3}$, and H$_{\beta}$ lines of IM~Peg (left panels)
                      and $\sigma$~Gem (right panels), respectively. The label identifying each observing cycle is marked in the top panels.}
      \label{fig2}
\end{figure*}
\begin{figure*}
    \centering
    \includegraphics[width=12.0cm,height=12.0cm, angle=270]{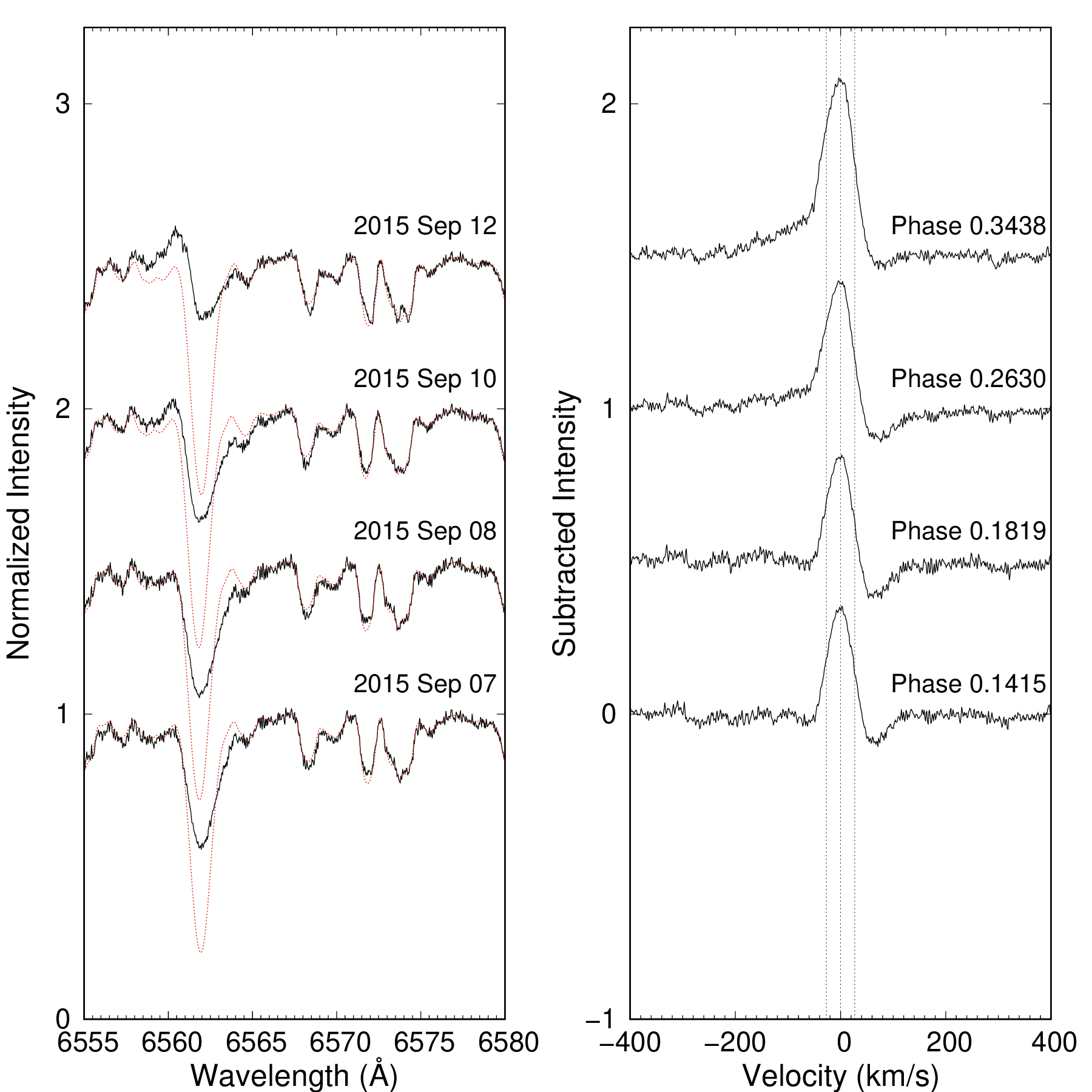}
     \caption{Series of the H$_{\alpha}$ line spectra of IM~Peg observed on 2015 September 07, 08, 10 and 12. The left panel are observed (solid lines) and synthesized (dotted lines) spectral, while the right one are subtracted spectra. For the right panel, the central dotted line marks the rest velocity of the stellar photosphere and the other two vertical lines indicate the maximum Doppler shifts associated the rotational velocity, $vsini$.}
     \label{fig3}
\end{figure*}
\begin{figure}
    \centering
     \includegraphics[width=15cm,height=8.5cm, angle=270]{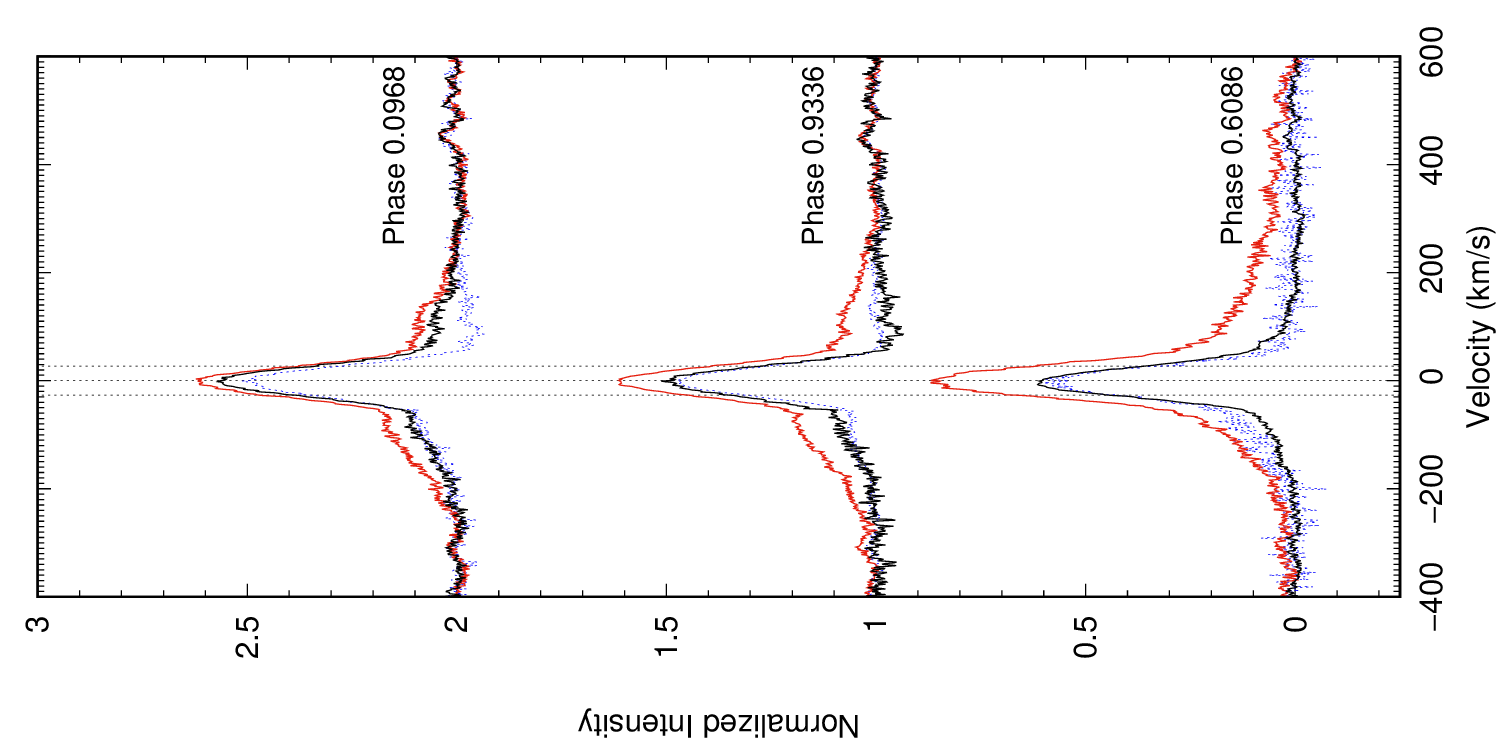}
     \caption{Subtracted H$_{\alpha}$ line profiles of IM~Peg at phases 0.6086, 0.9336 and 0.0968 during flare-like events (red solid lines), which are respectively overlapped 
                       with the former (blue dotted lines) and latter spectra (black solid lines) of each one. The central dotted line marks the rest velocity of the stellar photosphere and the 
                       other two vertical lines indicate the maximum Doppler shifts associated the rotational velocity, $vsini$.}
      \label{fig4}
\end{figure}
\begin{figure*}
    \centering
    \includegraphics[width=15cm,height=13cm, angle=270]{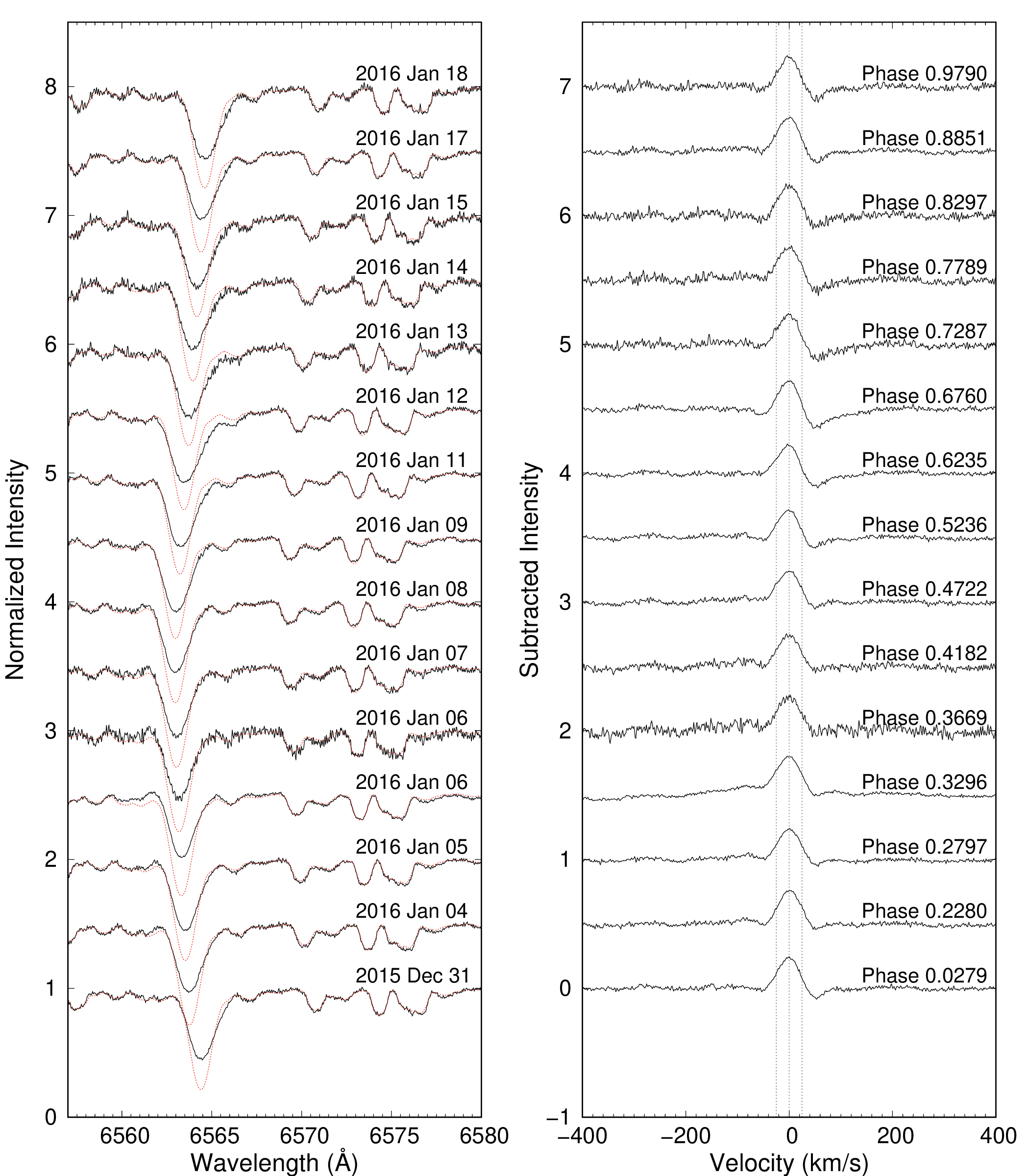}
     \caption{Series of the H$_{\alpha}$ line spectra of $\sigma$~Gem. The left panel are observed (solid lines) and synthesized (dotted lines) spectral, while the right one are subtracted spectra. For the right panel, the central dotted line marks the rest velocity of the stellar photosphere and the other two vertical lines indicate the maximum Doppler shifts associated the rotational velocity, $vsini$.}
     \label{fig5}
\end{figure*}
\section{Analysis of chromospheric activity indicators}
\indent
In this work we analyzed the chromospheric activity indicators H$_{\alpha}$, $\mbox{Na~{\sc i}}$ D$_{1}$, D$_{2}$ doublet, $\mbox{He~{\sc i}}$ D$_{3}$, and H$_{\beta}$. These lines are formed at different atmospheric heights with  different temperatures. To isolate the activity contribution from the photospheric absorption profiles in these activity lines, we used the synthesized spectra subtraction technique for all the observed spectra with the help of the STARMOD program \citep{barden1985, montes1995a, montes1995b, montes1997, montes2000}, which can simulate the photospheric absorption line profile (synthesized spectrum) by using the observed reference spectrum. Therefore, the subtraction between the observed and the synthesized spectrum provides the activity contribution as excess emission or absorption features above or below the continuum level. This technique has been widely used for chromospheric activity studies \citep[e.g.][]{montes1995a, montes1995b, montes1997, montes2000, gu2002, cao2015} and has also been applied for the detection of stellar prominences in binary systems \citep{Hall1992, cao2012, cao2019}. Since our target stars, IM~Peg and $\sigma$~Gem, are single-lined binary systems, we respectively used only one reference star spectrum to reproduce the synthesized spectrum.\\
\indent
For IM~Peg, we used the spectrum of the inactive star HR~5616 (K2~III) as reference for the primary in the synthesized spectrum construction. The rotational velocity ($vsini$) value was determined 
using the reference spectrum based on the method described in detail by \citet{barden1985}. The value of 27~km~s$^{-1}$ was obtained from high signal to noise ratio (SNR) spectra, spanning 
the wavelength regions 6370--6520~\AA~with many photospheric absorption lines, which is in good agreement with the results of 24.0~km~s$^{-1}$ estimated by \citet{Dempsey1992}, 25.6 $\pm$ 1~km~s$^{-1}$ by \citet{Medeiros1995}, 24~km~s$^{-1}$ by \citet{Ottmann1998}, 25 $\pm$ 1~km~s$^{-1}$ by \citet{Donati1997}, 28.2 $\pm$ 1~km~s$^{-1}$ by \citet{Fekel1997}, and 26.5 $\pm$ 0.5~km~s$^{-1}$ by \citet{berdyugina1999}. Consequently, the synthesized spectra were constructed by rotationaly broadening the reference spectrum to the $vsini$ value derived above and shifting along the radial-velocity axis, to get best fits. Then, the subtracted spectra between the observed and synthesized spectra were calculated for IM~Peg.\\
\indent
For the spectral analysis of $\sigma$~Gem, we used the spectrum of the inactive star HR~3264 (K1~III) as a reference spectrum. A $vsini$ value of 25~km~s$^{-1}$ was obtained using our reference spectrum, which is in good agreement with the values of 27.0 $\pm$ 1~km~s$^{-1}$ derived by \citet{Eaton1990}, 25.0~km~s$^{-1}$ by \citet{Dempsey1992}, 25.0~km~s$^{-1}$ by \citet{Strassmeier1993}, 27.0 $\pm$ 0.2~km~s$^{-1}$ by \citet{Duemmler1997}, 27.5 $\pm$ 1~km~s$^{-1}$ by \citet{Kovari2001}, and 24.8~km~s$^{-1}$ by \citet{Roettenbacher2017}. The synthesized spectra were constructed by using the above $vsini$ value and shifting along the radial-velocity axis, and therefore the subtraction were calculated for all spectra of $\sigma$~Gem.\\
\indent
Examples of the synthesized spectra subtraction in the chromospheric activity indicator H$_{\alpha}$, $\mbox{Na~{\sc i}}$ D$_{1}$, D$_{2}$ doublet, $\mbox{He~{\sc i}}$ D$_{3}$, and H$_{\beta}$ line regions of IM~Peg and  $\sigma$~Gem are respectively presented in Fig.~\ref{fig1}. The equivalent widths (EWs) of the subtracted H$_{\alpha}$, $\mbox{He~{\sc i}}$ D$_{3}$, and H$_{\beta}$ line profiles were measured using the SPLOT task in the IRAF package, and are listed in Table~\ref{tab1} along with their errors for IM~Peg and in Table~\ref{tab2} for $\sigma$~Gem, respectively. We determined the EWs by integrating over the subtraction, and additionally measured them using Gaussian function fit. The errors for the measured EWs were estimated by using the difference between the measurements of two methods, which should include errors arising from the synthetic spectrum used for the subtraction and the data quality of that particular observation. For some of our observations of $\sigma$~Gem, because the SNRs are very low in $\mbox{He~{\sc i}}$ D$_{3}$, and H$_{\beta}$ line line region, they were not measured in the analysis. For both of stars, we plot the EWs of H$_{\alpha}$, $\mbox{He~{\sc i}}$ D$_{3}$, and H$_{\beta}$ subtraction as a function of orbital phase in Fig.~\ref{fig2}, respectively.\\
\indent
Although the subtracted spectrum often shows obvious core emission feature in the H$_{\beta}$ line (see Fig.~\ref{fig1}), the extra-absorption wing could compensates its intensity and therefore their EWs going to be negative for most of our observations of IM~Peg. In Table~\ref{tab1}, due to the EWs of the subtracted H$_{\beta}$ line are positive for some of our observations of IM~Peg, we also give the ratio of excess emission $E_{H{\alpha}}/E_{H{\beta}}$ with the correction:
\begin{equation}
\frac{E_{H{\alpha}}}{E_{H{\beta}}} = \frac{EW(H_{\alpha})}{EW(H_{\beta})}*0.2444*2.512^{(B-R)}
\end{equation}
derived by \citet{Hall1992}, which takes into account the absolute flux density in these lines and the color difference in the components. For IM~Peg, the color index $B-R$~=~2.02 was used for the calculation.\\ 
\section{Results and discussion}
\subsection{Chromospheric activity behavior}
\subsubsection{IM~Peg}
\indent
During our observations the H$_{\alpha}$ line profile exhibit variability, showing filled-in absorption feature or moderate single-peaked emission in the blue wing of the absorption line or double-peaked emission profile above the continuum level. For the double-peaked emission profile, the blue shoulder is often stronger than the red one, which means that there is blue-shifted component in the chromospheric emission. Similar characteristics of H$_{\alpha}$ line profile have also been found in the other single-lined RS~CVn-type stars, such as VY~Ari and HK~Lac \citep{Bopp1980, Biazzo2006}. \citet{Bopp1980} interpreted the H$_{\alpha}$ line profile changing from filled-in absorption to double-peaked emission profile on HK~Lac obtained within a month as prominence-like feature which could give rise to a flare eruption. The H$_{\alpha}$ line always shows a filled-in variable profile, but a pure emission above the continuum was not observed by \citet{Biazzo2006}. Therefore, IM~Peg is probably in the stage of more active during our observations than that of \citet{Biazzo2006}.\\
\indent
After applying the spectral subtraction technique, as examples displayed in Fig.~\ref{fig1}, we can see that the H$_{\alpha}$, $\mbox{Na~{\sc i}}$ D$_{1}$, D$_{2}$ doublet, and H$_{\beta}$ lines show clear chromospheric emission feature in the subtraction, while the $\mbox{He~{\sc i}}$ D$_{3}$ line shows absorption feature compared to the synthesized spectrum. The $\mbox{He~{\sc i}}$ D$_{3}$ line in absorption has also been found on IM~Peg, VY~Ari and HK~Lac by \citet{Biazzo2006}, and in several other RS~CVn-type systems \citep{Wolff1984, montes1997, montes2000, cao2020}. Moreover, the subtracted H$_{\alpha}$ line profile has a broad blue-shifted emission component for most of time and an excess absorption sometimes which appears always red-shifted compared to the line center, as examples plotted in Fig.~\ref{fig3}, in which we display time-series of the observed and subtracted H$_{\alpha}$ line profiles obtained on 2015 September 07, 08, 10 and 12. For the H$_{\beta}$ line subtraction, there is a strong absorption wing for most of our observations. And the absorption wing is usually red asymmetric. Furthermore, because the $\mbox{Na~{\sc i}}$ D$_{1}$, D$_{2}$ doublet lines are more sensitive to the effective temperature, just light temperature difference between the target and reference stars can produce significant changes in the wings of the line profiles \citep{montes1997}. Therefore, the synthesized spectrum does not match the observational one well in the the $\mbox{Na~{\sc i}}$ D$_{1}$, D$_{2}$ doublet line regions for our situation.\\ 
\indent
The $E_{H{\alpha}}$/$E_{H{\beta}}$ values have been usually used as a diagnostic for discriminating the presence of different structures on the stellar surface. As \cite{Huenemoerder1987} discussed, the low ratios in RS CVn-type stars are caused by plage-like regions, while prominence-like structures have high values. Moreover, similar results have also been derived by \citet{Hall1992} who found that low ratios ($\sim$ 1-2) can be achieved both in plages and prominences viewed against the disk, but high values ($\sim$ 3-15) can only be obtained in extended structures viewed off the stellar limb. For our situation, the derived ratio values are very high ($\sim$ 3-230, see Table~\ref{tab1}), which are probably because the excess absorption could compensate the intensity of the subtraction on the one hand, especially the H$_{\beta}$ line. On the other hand, the high values mean that there are probably extended regions on IM~Peg, and more probably these extended structures resulted in excess absorption features seen in the H$_{\alpha}$ and H$_{\beta}$ subtraction. \\
\indent
For our situation, we have a long-term continuous observations which covered five orbital cycles, and therefore we group the observations of each cycle together. From Fig.~\ref{fig2}, it can be seen that the EW variation of the H$_{\alpha}$, $\mbox{He~{\sc i}}$ D$_{3}$, and H$_{\beta}$ line subtractions correlates well and shows rotational modulation. When the intensity becomes much stronger in the H$_{\alpha}$ and H$_{\beta}$ lines, the $\mbox{He~{\sc i}}$ D$_{3}$ line absorption becomes much lower. Similar correlation between the $\mbox{He~{\sc i}}$ D$_{3}$ and H$_{\alpha}$ line subtractions has also been found by \citet{Biazzo2006} on IM~Peg and VY~Ari. For the different orbital cycle of IM~Peg, the EWs of chromospheric activity indicators shows different variation trend, which implies the evolution of activity regions. For Cycle~1, the activity level is much flat between phases 0.2 and 0.8, but relatively lower in the rest of phase coverage because there are red-shifted excess absorption feature in the H$_{\alpha}$ line subtraction. In Cycle~2, the activity variation is similar to Cycle~1, except for two sudden points which have much stronger EWs, probably caused by flare-like events. Moreover, another weaker sudden point could also be found in Cycle~3, in which the phase coverage is not good. In Cycles~4 and 5, it is obvious that there is an active longitude near phases 0.45--0.5. Furthermore, there are excess absorption features observed near phases 0.8--0.9 for Cycle~4 and phases 0.1--0.3 for Cycle~5. On the average, therefore, the red-shifted excess absorption was usually observed at phases from 0.8 to 0.3 during our observations.\\
\indent
The $\mbox{He~{\sc i}}$ D$_{3}$ line is indicative of optical flares due to its very high excitation potential. When an strong optical flare happened, the $\mbox{He~{\sc i}}$ D$_{3}$ line shows obvious emission feature, which has been widely observed during flares in solar \citep{Zirin1988} and stellar chromospheres, especially in RS~CVn-type stars (e.g., \citealt{cao2017, cao2019}). Although we have not detected emission feature in the $\mbox{He~{\sc i}}$ D$_{3}$ line in our spectra, there are three EW points which are much stronger in the H$_{\alpha}$ and H$_{\beta}$ line subtractions, and meanwhile the $\mbox{He~{\sc i}}$ D$_{3}$ line shows a strong filled-in absorption. The first and second ones are respectively at phases 0.6086 and 0.9336 in Cycle~2 and the other weaker one is at phase 0.0968 in Cycle~3. Moreover, these stronger EW points are suddenly taken place when compared to the former and latter of each one, as shown in Fig.~\ref{fig4}, in which we plot the subtracted H$_{\alpha}$ spectra during flare-like events. Therefore, according to these observing facts, we interpret that these observations are probably caused by flare-like events, especially the first one because it has the most intense EW during our observations. For the first flare-like event, we calculated the stellar continuum flux $F_{H_{\alpha}}$~(erg cm$^{-2}$ s$^{-1}$ \AA$^{-1}$) in the H$_{\alpha}$ line region as a function of the color index $B-V$= 1.12 based on the empirical relationship
\begin{eqnarray}
\log{F_{H_{\alpha}}}=[7.538-1.081(B-V)]\pm{0.33}\nonumber \\
0.0~\leq~B-V~\leq~1.4
\end{eqnarray} 
of \citet{hall1996}, and then converted the EW into the absolute flux at the stellar surface $F_{S}$~(erg~cm$^{-2}$~s$^{-1}$). Therefore, the flare energy in the observed H$_{\alpha}$ line, $6.7 \times 10^{31}$ erg~s$^{-1}$, was derived through converting the absolute surface fluxes into luminosity by using the radius R$_{\ast}$~=~13.3~R$_{\sun}$ of the K2~III primary star \citep{berdyugina1999}. The value for the energy released in the H$_{\alpha}$ line is of similar order of magnitude as the strong optical flares detected in the other very active RS CVn-type stars like UX~Ari \citep{montes1996, gu2002} and SZ~Psc \citep{cao2019, cao2020}.\\
\subsubsection{$\sigma$~Gem}
\indent
As examples shown in Fig.~\ref{fig1}, chromospheric activity indicator H$_{\alpha}$, $\mbox{Na~{\sc i}}$ D$_{1}$, D$_{2}$ doublet, $\mbox{He~{\sc i}}$ D$_{3}$, and H$_{\beta}$ line profiles of $\sigma$~Gem are always characterized by absorption in our spectra. Especially the observed H$_{\alpha}$ line profile, there was no pure emission feature above the continuum level, which is in good agreement with the previous observations (e.g., \citealt{Strassmeier1990, Zhang1999}). Therefore, $\sigma$~Gem shows an activity level lower than that of IM~Peg.\\
\indent
After applying the spectral subtraction technique, shown in Fig.~\ref{fig1}, it can be seen that the H$_{\alpha}$, $\mbox{Na~{\sc i}}$ D$_{1}$, D$_{2}$ doublet, and H$_{\beta}$ lines show a similar behavior as the star IM~Peg. The subtracted H$_{\alpha}$ line shows obvious chromospheric emission feature, but with an excess absorption wing in most of our spectra. As the sample of the observed and subtracted H$_{\alpha}$ line profiles plotted in Fig.~\ref{fig5}, the absorption feature usually appears red-shifted compared to the line center. For some of our observations, moreover, the H$_{\alpha}$ line subtraction has also the blue-shifted emission wing, which appears less frequently and is much weaker in comparison with IM~Peg. A much stronger blue-shifted emission wing was found at phases 0.3296 and 0.3297 on 2016 Jan~6. Furthermore, the H$_{\beta}$ line subtraction shows an obvious core emission, also with a strong absorption wing. The $\mbox{He~{\sc i}}$ D$_{3}$ line is also in absorption in all of our spectra. In addition, the synthesized spectrum of the $\mbox{Na~{\sc i}}$ D$_{1}$, D$_{2}$ doublet shows the same situation as IM~Peg.\\
\indent
During our long-term observations, there are eight orbital cycles covered in the observing run, and therefore we group the observations of each cycle together. Same as the activity behavior of IM~Peg, showing in Fig.~\ref{fig2}, it can be see that the EW variation of the H$_{\alpha}$, $\mbox{He~{\sc i}}$ D$_{3}$, and H$_{\beta}$ line subtraction of $\sigma$~Gem correlates well and shows different behavior in different cycles. The EWs have a good rotational modulation in the second half of the orbital phase among different cycles and there is a possibly narrow activity longitude around phases 0.8--0.9, but show relative scatter in the first half of the orbital phase. In the first half of the orbital phase, there are different activity longitudes appeared in different cycles: around phase 0.25 in Cycle~1, phases 0.3--0.5 in Cycle~2, near phase 0.3 in Cycle~4, around phase 0.25 in Cycle~5. In Cycles 3, 6, 7 and 8, there are no prominent activity longitudes in the first half of the orbital phase.\\
\indent
Using the same data as us, \citet{Korhonen2021} reconstructed surface temperature maps with Doppler imaging technique and analyzed the variability of the H$_{\alpha}$ line. In their Doppler maps, the high-latitude or polar spots concentrated around phase 0.25 at the beginning of the observing run, which are probably spatially associated with the chromospheric activity longitudes in the first half of the orbital phase derived by us, and then moved to phase 0.75 by the end of the observations. Moreover, to enhance the  H$_{\alpha}$ line variation caused by chromospheric emission, an average line profile constructed from all observed spectra was subtracted from the individual H$_{\alpha}$ line profile by \citet{Korhonen2021}. Through comparing average photospheric temperature with summed H$_{\alpha}$ subtraction, \citet{Korhonen2021} also derived a weak correlation between the locations of starspots and enhanced chromospheric emission at some epochs. Different from our results, \citet{Korhonen2021} just found the excess absorption features at some phases around the spectral region of H$_{\alpha}$ line. For their method, the excess absorption features would appeared when the H$_{\alpha}$ line profile and/or intensity occurred variation. However, based on the synthesized spectral subtraction technique in our analysis, the excess absorption is a local feature and usually
appears red-shifted in most of our subtracted spectra (see examples shown in Fig.~\ref{fig5}). \\
\subsection{Asymmetric H$_{\alpha}$ subtraction}
\indent
For our observations, it is worth to noticing that there are excess absorption features in the H$_{\alpha}$ line subtraction of IM~Peg and $\sigma$~Gem, as examples shown in Figs.~\ref{fig4} and \ref{fig5}, respectively. The excess absorption features could not result from systematic problems in the template spectra that are used for the synthesized spectra subtraction. Because they are evolved features and observed by the other author. For example, the excess absorption features usually appeared red-shifted compared to the H$_{\alpha}$ line center for both stars and evolved with time. And for the star $\sigma$~Gem, there is an exception that a strong blue-shifted excess absorption feature was found at phases 0.4264, 0.4265, and 0.4266 on 2016 Mar 6. \citet{Biazzo2006} have found same red-shifted excess absorption feature in the H$_{\alpha}$ line subtraction on IM~Peg. Moreover, similar excess absorption feature, especially in the H$_{\alpha}$ line, has been sometimes found in several RS CVn-type stars which mainly are double-lined eclipsing binaries and usually seen just before and during the primary eclipse \citep{Hall1990, Hall1992, Frasca2000, cao2019, cao2020}. These excess absorption features have been interpreted as caused by prominence-like materials which could project against a significant fraction of the stellar disk, and therefore scatter the underlying chromospheric emission out of the line of sight or absorb the continuum from the star, especially from the secondary component of the binary system. In single rapid rotating late-type stars like active K dwarfs AB Dor (e.g., \citealt{Collier1989a, Collier1989b}) and HD~197890 (e.g., \citealt{Dunstone2006}), prominences have also been detected as transient absorption features moved rapidly across the rotationally broadened H$_{\alpha}$ line profile from blue edge to red .\\
\indent
However, our target stars are single-lined binaries and more importantly the excess absorption feature always appears red-shifted compared to the line center, especially on the star IM~Peg (see Fig.~\ref{fig4}). For the single rapid rotator BD+22$\degr$4409, \citet{Eibe1999} found the observed H$_{\alpha}$ line profile is always very asymmetric because of a strong narrower absorption feature at positive velocities and therefore interpreted as due to an intense and continuous downflow of absorbing material, which is more likely to be produced as a result of the interaction between the magnetic field and the stellar plasma in conditions of unstable mechanical equilibrium, resembling the loop prominence phenomenon. Therefore, the red-shifted excess absorption features in the  H$_{\alpha}$ line subtraction of our two target stars could also be interpreted as strong down-flow of cool absorbing materials.\\
\indent
Moreover, one other activity phenomenon of concern is that there is a broad blue-shifted emission component in the H$_{\alpha}$ line subtraction, sometimes very strong in the spectra of IM~Peg. The broad emission wings have been usually observed in several RS~CVn-type binary stars, and interpreted as arising from microflaring by \citet{montes1997}. Microflares are the low-energy extension of flares, and therefore have large-scale material motions which could explain the broad wings observed in the H$_{\alpha}$ and other chromospheric activity lines. The material motions along the line of sight could result in the broad wings in spectral lines. Therefore, for our situation, the blue-shifted broad emission component in the H$_{\alpha}$ line wing could originate from up-flow of hot materials during microflare-like events. And these up-flow materials are probably the original source of intense down-flow of cool absorbing materials.\\
\section{Summary and conclusions}
\indent
Based on the above analysis of continuous high-resolution spectroscopic observations taken during the 2015--2016 season, we have obtained information about chromospheric activity of two long-period single-lined RS~CVn-type stars IM~Peg and $\sigma$~Gem with several chromospheric activity indicators. The following main results are obtained.\\
\begin{enumerate}
\item
Chromospheric emission in the H$_{\alpha}$, $\mbox{Na~{\sc i}}$ D$_{1}$, D$_{2}$ doublet, and H$_{\beta}$ lines confirms that IM~Peg and $\sigma$~Gem are both of very active stars. On the average, the activity level of IM~Peg is much stronger than that of $\sigma$~Gem. Moreover, the $\mbox{He~{\sc i}}$ D$_{3}$ line is always in absorption in our spectra for both stars. In addition, flare-like events are detected on IM~Peg, during which the chromospheric emission suddenly increases.
\item 
The EW variation of H$_{\alpha}$, $\mbox{He~{\sc i}}$ D$_{3}$, and H$_{\beta}$ line subtraction correlates well for both stars. When the subtraction intensity becomes much stronger in the H$_{\alpha}$ and H$_{\beta}$ lines, the $\mbox{He~{\sc i}}$ D$_{3}$ line has a shallower absorption. Moreover, the EWs presents different variation behavior among different cycles and shows rotation modulation, which suggests the presence and evolution of the chromospheric active longitudes over the surface of both stars during our observations.
\item 
The subtraction, especially the H$_{\alpha}$ line, shows interesting asymmetric feature. The red-shifted excess absorption features could be interpreted as strong down-flow of cool absorbing materials, while the blue-shifted emission component is probably caused by up-flow of hot materials through microflare events.
\end{enumerate}
\section*{Acknowledgements}
Based on observations made with the Hertzsprung SONG telescope operated on the Spanish Observatorio del Teide on the island of Tenerife by the Aarhus and Copenhagen Universities and by the Instituto de Astrof\'{i}sica de Canarias. Funding for the Stellar Astrophysics Centre is provided by The Danish National Research Foundation (Grant agreement no. DNRF106). This work also is financially supported by the National Natural Science Foundation of China (NSFC) under grants Nos. 10373023, 10773027, 11333006, U1531121, and 11903074, and the Chinese Academy of Sciences (CAS) ``Light of West China'' Program. We also acknowledge the science research grant from the China Manned Space Project with NO. CMS-CSST-2021-B07. We would also like to thank the anonymous referee for helpful comments and suggestions, which significantly improve our manuscript.
\section*{Data availability}
The data underlying this article are available from the SONG Data Archive (SODA) or from the author upon request. 



\begin{thebibliography}{99}
\bibitem[\protect\citeauthoryear{Andersen et al.}{2014}]{Andersen2014} Andersen M. F., Grundahl F., Christensen-Dalsgaard J., et al. 2014, RMxAC, 45, 83
\bibitem[\protect\citeauthoryear{Andersen et al.}{2019}]{Andersen2019} Andersen, M. F., Handberg, R., Weiss, E., et al. 2019, PASP, 131, 045003
\bibitem[\protect\citeauthoryear{Antoci et al.}{2013}]{Antoci2013} Antoci V., Handler G., Grundahl F., et al. 2013, \mnras, 435, 1563
\bibitem[\protect\citeauthoryear{Ayres, Simon, \& Linsky}{1984}]{Ayres1984} Ayres T. R., Simon T., Linsky J. L., 1984, \apj, 279, 197
\bibitem[\protect\citeauthoryear{Barden}{1985}]{barden1985} Barden S. C., 1985, \apj, 295, 162
\bibitem[\protect\citeauthoryear{Berdyugina et al}{2000}]{berdyugina2000} Berdyugina S. V., Berdyugina A. V., Ilyin I., Tuominen I., 2000, \aap, 360, 272
\bibitem[\protect\citeauthoryear{Berdyugina, Ilyin, \& Tuominen}{1999}]{berdyugina1999} Berdyugina S. V., Ilyin I., Tuominen I., 1999, \aap, 347, 932
\bibitem[\protect\citeauthoryear{Biazzo et al.}{2006}]{Biazzo2006} Biazzo K., Frasca A., Catalano S., Marilli E., 2006, \aap, 446, 1129
\bibitem[\protect\citeauthoryear{Bopp \& Talcott}{1980}]{Bopp1980} Bopp B. W., Talcott J. C., 1980, \aj, 85, 55
\bibitem[\protect\citeauthoryear{Cao \& Gu}{2012}]{cao2012} Cao D. T., Gu S. H., 2012, \aap, 538, 130
\bibitem[\protect\citeauthoryear{Cao \& Gu}{2015}]{cao2015} Cao D. T., Gu S. H., 2015, \mnras, 449, 1380
\bibitem[\protect\citeauthoryear{Cao \& Gu}{2017}]{cao2017} Cao D. T., Gu S. H., 2017, RAA, 17, 055
\bibitem[\protect\citeauthoryear{Cao et al.}{2019}]{cao2019} Cao D. T., Gu S. H., Ge J., et al. 2019, \mnras, 482, 988
\bibitem[\protect\citeauthoryear{Cao et al.}{2020}]{cao2020} Cao D. T., Gu S. H., Wolter U., et al. 2020, \aj
\bibitem[\protect\citeauthoryear{Collier Cameron \& Robinson}{1989a}]{Collier1989a} Collier Cameron A., Robinson R. D., 1989a, \mnras, 236, 57
\bibitem[\protect\citeauthoryear{Collier Cameron \& Robinson}{1989b}]{Collier1989b} Collier Cameron A., Robinson R. D., 1989b, \mnras, 238, 657
\bibitem[\protect\citeauthoryear{Corsaro et al.}{2012}]{Corsaro2012} Corsaro E., Grundahl F., Leccia S., et al. 2012, \aap, 537, A9
\bibitem[\protect\citeauthoryear{De Medeiros \& Mayor}{1995}]{Medeiros1995} De Medeiros J. R., Mayor M., 1995, \aap, 302, 745
\bibitem[\protect\citeauthoryear{Dempsey et al.}{1993}]{Dempsey1993} Dempsey R. C., Bopp B. W., Henry G. W., Hall D. S., 1993, \apjs, 86, 293
\bibitem[\protect\citeauthoryear{Dempsey et al.}{1992}]{Dempsey1992} Dempsey R. C., Bopp B. W., Strassmeier K. G., et al. 1992, \apj, 392, 187 
\bibitem[\protect\citeauthoryear{Dempsey et al.}{1996}]{Dempsey1996}  Dempsey R. C., Neff J. E., O'Neal D., Olah K., 1996, \aj, 111, 1356
\bibitem[\protect\citeauthoryear{Donati et al.}{1997}]{Donati1997} Donati J,-F., Semel M., Carter B. D., Rees D. E., Cameron A. C., 1997, \mnras, 291, 658
\bibitem[\protect\citeauthoryear{Duemmler, Ilyin, \& Tuominen}{1997}]{Duemmler1997} Duemmler R., Ilyin I. V., Tuominen I., 1997, \aaps, 123, 209D 
\bibitem[\protect\citeauthoryear{Dunstone et al.}{2006}]{Dunstone2006} Dunstone N. J., Barnes J. R., Collier Cameron A., Jardine M., 2006, \mnras, 365, 530 
\bibitem[\protect\citeauthoryear{Eaton}{1990}]{Eaton1990} Eaton J. A., 1990, IBVS 3460
\bibitem[\protect\citeauthoryear{Eibe et al.}{1999}]{Eibe1999} Eibe M. T., Byrne P. B., Jeffries R. D., Gunn A. G., 1999, \aj, 341, 527
\bibitem[\protect\citeauthoryear{Fekel}{1997}]{Fekel1997} Fekel F. C., 1997, PASP, 109, 154
\bibitem[\protect\citeauthoryear{Frasca \& Catalano}{1994}]{Frasca1994} Frasca A., Catalano S., 1994, \aap, 284, 883
\bibitem[\protect\citeauthoryear{Frasca et al.}{2000}]{Frasca2000} Frasca A., Marino G., Catalano S., Marilli E., 2000, \aap, 358, 1007
\bibitem[\protect\citeauthoryear{Grundahl et al.}{2017}]{Grundahl2017} Grundahl F., Andersen M. F., Christensen-Dalsgaard J., et al. 2017, ApJ, 836, 142
\bibitem[\protect\citeauthoryear{Grundahl et al.}{2009}]{Grundahl2009} Grundahl F., Christensen-Dalsgaard J., Kjeldsen H., et al. 2009, in Astronomical Society of the Pacific Conference Series, Vol. 416, Solar-Stellar Dynamos as Revealed by Helio- and Asteroseismology: GONG 2008/SOHO 21, ed. M. Dikpati, T. Arentoft, I. Gonz\'{a}lez Hern\'{a}ndez, C. Lindsey, \& F. Hill, 579
\bibitem[\protect\citeauthoryear{Grundahl et al.}{2006}]{Grundahl2006} Grundahl F., Kjeldsen H., Frandsen S., et al. 2006, Mem. Soc. Astron. Italiana, 77, 458
\bibitem[\protect\citeauthoryear{Gu et al.}{2002}]{gu2002} Gu S.-H., Tan H.-S., Shan H.-G., Zhang F.-H., 2002, A\&A, 388, 889
\bibitem[\protect\citeauthoryear{Hall}{1976}]{hall1976} Hall, D. S. 1976, in Fitch W. S., ed, Multiple Periodic Variable Stars, IAU Coll. 29. Reidel: Dordrecht, p. 287
\bibitem[\protect\citeauthoryear{Hall, Henry, \& Landis}{1977}]{Hall1977} Hall D. S., Henry G. W., Landis H. W., 1977, IBVS, 1328, 1
\bibitem[\protect\citeauthoryear{Hall}{1996}]{hall1996} Hall J. C., 1996, PASP, 108, 313
\bibitem[\protect\citeauthoryear{Hall \& Ramsey}{1992}]{Hall1992} Hall J. C., Ramsey L. W., 1992, \aj, 104, 142
\bibitem[\protect\citeauthoryear{Hall et al.}{1990}]{Hall1990} Hall, J. C., Huenemoerder D. P., Ramsey, L. W., Buzasi D. L., 1990, \apj, 358, 610
\bibitem[\protect\citeauthoryear{Hatzes}{1993}]{Hatzes1993} Hatzes A. P., 1993, \apj, 410, 777
\bibitem[\protect\citeauthoryear{Huenemoerder \& Ramsey}{1987}]{Huenemoerder1987} Huenemoerder D. P., Ramsey L. W., 1987, \apj, 319, 392
\bibitem[\protect\citeauthoryear{Huenemoerder, Ramsey, \& Buzasi}{1990}]{Huenemoerder1990} Huenemoerder D. P., Ramsey L. W., Buzasi D. L., 1990, \apj, 350, 763
\bibitem[\protect\citeauthoryear{Korhonen et al.}{2021}]{Korhonen2021} Korhonen H., Roettenbacher R. M., Gu S., et al. 2021, \aap, 646, A6
\bibitem[\protect\citeauthoryear{Kov\'{a}ri et al.}{2001}]{Kovari2001} Kov\'{a}ri Zs., Strassmeier K. G., Bartus J., Washuettl A., Weber M., Rice J. B., 2001, \aap, 373, 199
\bibitem[\protect\citeauthoryear{Landman \& Mongillo}{1979}]{Landman1979} Landman D. A., Mongillo M., 1979, \apj, 230, 581
\bibitem[\protect\citeauthoryear{Marsden et al.}{2005}]{Marsden2005} Marsden S. C., Berdyugina S. V., Donati J.-F.,  et al., 2005, \apj, 634, L173
\bibitem[\protect\citeauthoryear{Montes et al.}{1995a}]{montes1995a} Montes D., Fern\'{a}ndez-Figueroa M. J., De Castro E., Cornide M., 1995a, A\&A, 294, 165
\bibitem[\protect\citeauthoryear{Montes et al.}{1995b}]{montes1995b} Montes D., Fern\'{a}ndez-Figueroa M. J., De Castro E., Cornide M., 1995b, A\&AS, 109, 135
\bibitem[\protect\citeauthoryear{Montes et al.}{2000}]{montes2000} Montes D., Fern\'{a}ndez-Figueroa M. J., De Castro E., Cornide M., Latorre A., Sanz-Forcada J., 2000, A\&AS, 146, 103
\bibitem[\protect\citeauthoryear{Montes et al.}{1997}]{montes1997} Montes D., Fern\'{a}ndez-Figueroa M. J., De Castro E., Sanz-Forcada J., 1997, A\&AS, 125, 263
\bibitem[\protect\citeauthoryear{Montes et al.}{1996}]{montes1996} Montes D., Sanz-Forcada J., Fern\'{a}ndez-Figueroa M. J., Lorente R., 1996, \aap, 310, L29
\bibitem[\protect\citeauthoryear{Ottmann, Pfeiffer, \& Gehren}{1998}]{Ottmann1998} Ottmann R., Pfeiffer M. J., Gehren T., 1998, A\&A, 338, 661
\bibitem[\protect\citeauthoryear{Roettenbacher et al.}{2015}]{Roettenbacher2015} Roettenbacher R. M., Monnier J. D., Henry G. W., et al. 2015, \apj, 807, 23
\bibitem[\protect\citeauthoryear{Roettenbacher et al.}{2017}]{Roettenbacher2017} Roettenbacher R. M., Monnier J. D., Korhonen H., et al. 2017, \apj, 849, 120
\bibitem[\protect\citeauthoryear{Schrijver \& Zwaan}{2000}]{schrijver2000} Schrijver C. J., Zwaan C., 2000, Solar and stellar magnetic activity. Cambridge Univ. Press, Cambridge
\bibitem[\protect\citeauthoryear{Strassmeier et al.}{1997}]{Strassmeier1997} Strassmeier K.G., Bartus J., Cutispoto G., Rodon\'{o} M., 1997, \aaps, 125, 11
\bibitem[\protect\citeauthoryear{Strassmeier et al.}{1990}]{Strassmeier1990} Strassmeier K. G., Fekel F. C., Bopp B. W., Dempsey R. C., Henry G. W., 1990, \apjs, 72, 191
\bibitem[\protect\citeauthoryear{Strassmeier et al.}{1993}]{Strassmeier1993} Strassmeier K. G., Hall D. S., Fekel F. C., Scheck M., 1993, \aaps, 100, 173
\bibitem[\protect\citeauthoryear{Uytterhoeven et al.}{2012}]{Uytterhoeven2012} Uytterhoeven K.,  Pall\'{e} P. L., Grundahl F., et al. 2012, AN, 333, 1103
\bibitem[\protect\citeauthoryear{Wolff \& Heasley}{1984}]{Wolff1984} Wolff S. C., Heasley J. N., 1984, Pub. ASP, 96, 238
\bibitem[\protect\citeauthoryear{Zhang \& Zhang}{1999}]{Zhang1999} Zhang X. B., Zhang R. X., 1999, \aaps, 137, 217
\bibitem[\protect\citeauthoryear{Zirin}{1988}]{Zirin1988} Zirin H., 1988, Astrophysics of the Sun. Cambridge Univ. Press, Cambridge
\end{thebibliography}



\appendix
\section{Observing logs and measurements for the H$_{\alpha}$, $\mbox{He~{\sc i}}$ D$_{3}$, and H$_{\beta}$ line subtraction.}
\begin{table*}
\centering
\caption{Observing log and measurements for the subtraction of IM~Peg.}
\tabcolsep 0.25cm
\label{tab1}
\begin{tabular}{@{}cccccccc}
\hline
\bf{Date} & \bf{HJD} & \bf{Phase} & \bf{Cycle} &\multicolumn{3}{c}{\bf{EW({\AA})}} & $\frac{\bf{E_{H{\alpha}}}}{\bf{E_{H{\beta}}}}$\\
\cline{5-7}\\
 & (2,457,000+) & & & H$_{\alpha}$ & $\mbox{He~{\sc i}}$ D$_{3}$ & H$_{\beta}$\\
\hline
   2015/05/06&148.6845 &0.1103&1&0.378$\pm$0.019&-0.118$\pm$0.005&-0.184$\pm$0.015&...\\
   2015/05/07&149.6827 &0.1508&1&0.617$\pm$0.012&-0.107$\pm$0.007&-0.118$\pm$0.011&...\\
   2015/05/08&150.6835 &0.1914&1&0.879$\pm$0.012&-0.100$\pm$0.006&-0.049$\pm$0.017&...\\
   2015/05/09&151.7081 &0.2330&1&0.694$\pm$0.011&-0.115$\pm$0.008&-0.156$\pm$0.011&...\\
   2015/05/10&152.7019 &0.2733&1&0.837$\pm$0.014&-0.102$\pm$0.008&-0.036$\pm$0.012&...\\
   2015/05/12&154.6889 &0.3539&1&0.863$\pm$0.015&-0.088$\pm$0.006&-0.032$\pm$0.012&...\\
   2015/05/13&155.6956 &0.3948&1&0.803$\pm$0.012&-0.111$\pm$0.007&-0.029$\pm$0.012&...\\
   2015/05/17&159.6896 &0.5568&1&1.013$\pm$0.025&-0.082$\pm$0.004&0.007$\pm$0.012&227.33\\
   2015/05/18&160.6911 &0.5974&1&1.079$\pm$0.012&-0.092$\pm$0.004&0.026$\pm$0.012&65.19\\
   2015/05/19&161.6915 &0.6380&1&0.881$\pm$0.025&-0.100$\pm$0.006&-0.030$\pm$0.013&...\\
   2015/05/21&163.7049 &0.7197&1&0.807$\pm$0.018&-0.109$\pm$0.005&-0.058$\pm$0.011&...\\
   2015/05/22&164.6496 &0.7580&1&1.100$\pm$0.027&-0.127$\pm$0.007&0.011$\pm$0.012&157.09\\
   2015/05/23&165.6557 &0.7988&1&0.811$\pm$0.026&-0.105$\pm$0.005&-0.015$\pm$0.015&...\\
   2015/05/24&166.6722 &0.8401&1&0.687$\pm$0.013&-0.113$\pm$0.006&-0.109$\pm$0.016&...\\
   2015/05/25&167.7002 &0.8818&1&0.562$\pm$0.011&-0.101$\pm$0.005&-0.143$\pm$0.015&...\\
   2015/05/26&168.6135 &0.9188&1&0.543$\pm$0.021&-0.111$\pm$0.006&-0.119$\pm$0.011&...\\
   2015/05/27&169.6485 &0.9608&1&0.547$\pm$0.013&-0.096$\pm$0.008&-0.110$\pm$0.012&...\\
   2015/05/29&171.7099 &0.0444&2&0.853$\pm$0.026&-0.101$\pm$0.006&-0.055$\pm$0.012&...\\
   2015/05/30&172.6411 &0.0822&2&0.936$\pm$0.024&-0.098$\pm$0.005&0.013$\pm$0.015&113.10\\
   2015/06/01&174.6117 &0.1622&2&0.866$\pm$0.018&-0.103$\pm$0.007&-0.080$\pm$0.015&...\\
   2015/06/03&176.6108 &0.2433&2&0.723$\pm$0.011&-0.120$\pm$0.006&-0.053$\pm$0.016&...\\
   2015/06/09&182.6506 &0.4883&2&1.188$\pm$0.014&-0.090$\pm$0.008&0.039$\pm$0.015&47.85\\
   2015/06/10&183.5943 &0.5266&2&1.163$\pm$0.012&-0.099$\pm$0.005&0.022$\pm$0.015&83.04\\
   2015/06/11&184.6372 &0.5689&2&1.166$\pm$0.013&-0.048$\pm$0.008&0.077$\pm$0.016&23.79\\
   2015/06/12&185.6156 &0.6086&2&2.925$\pm$0.024&-0.031$\pm$0.008&1.128$\pm$0.037&4.07\\
   2015/06/15&188.7225 &0.7347&2&1.148$\pm$0.018&-0.093$\pm$0.005&0.071$\pm$0.012&25.40\\
   2015/06/16&189.6536 &0.7724&2&1.047$\pm$0.014&-0.094$\pm$0.005&0.049$\pm$0.011&33.57\\
   2015/06/17&190.6287 &0.8120&2&0.879$\pm$0.014&-0.118$\pm$0.009&-0.095$\pm$0.012&...\\
   2015/06/18&191.5851 &0.8508&2&0.758$\pm$0.015&-0.127$\pm$0.008&-0.088$\pm$0.013&...\\
   2015/06/19&192.6288 &0.8931&2&0.769$\pm$0.013&-0.106$\pm$0.006&-0.063$\pm$0.012&...\\
   2015/06/20&193.6274 &0.9336&2&1.749$\pm$0.026&-0.033$\pm$0.009&0.712$\pm$0.012&3.86\\
   2015/06/21&194.6023 &0.9732&2&0.815$\pm$0.019&-0.096$\pm$0.008&-0.005$\pm$0.011&...\\
   2015/06/24&197.6487 &0.0968&3&1.618$\pm$0.012&-0.081$\pm$0.005&0.083$\pm$0.012&30.62\\
   2015/06/25&198.6077 &0.1357&3&1.186$\pm$0.015&-0.109$\pm$0.007&-0.019$\pm$0.014&...\\
   2015/06/26&199.6279 &0.1771&3&0.897$\pm$0.014&-0.114$\pm$0.006&-0.070$\pm$0.012&...\\
   2015/06/27&200.6517 &0.2186&3&0.775$\pm$0.011&-0.111$\pm$0.007&-0.073$\pm$0.016&...\\
   2015/06/28&201.6506 &0.2591&3&0.760$\pm$0.026&-0.088$\pm$0.012&-0.094$\pm$0.012&...\\
   2015/06/30&203.6314 &0.3395&3&0.878$\pm$0.011&-0.073$\pm$0.009&-0.095$\pm$0.013&...\\
   2015/07/01&204.6247 &0.3798&3&1.145$\pm$0.017&-0.066$\pm$0.005&0.026$\pm$0.012&69.18\\
   2015/07/25&228.7174 &0.3572&4&1.285$\pm$0.024&-0.092$\pm$0.006&0.071$\pm$0.012&28.43\\
   2015/07/25&228.7265 &0.3576&4&1.280$\pm$0.024&-0.095$\pm$0.005&0.067$\pm$0.012&30.01\\
   2015/07/26&229.7313 &0.3984&4&1.635$\pm$0.011&-0.082$\pm$0.007&0.133$\pm$0.016&19.31\\
   2015/07/27&230.7015 &0.4377&4&2.298$\pm$0.021&-0.087$\pm$0.006&0.217$\pm$0.015&16.63\\
   2015/07/28&231.7017 &0.4783&4&2.374$\pm$0.020&-0.069$\pm$0.005&0.202$\pm$0.012&18.46\\
   2015/07/29&232.7112 &0.5193&4&1.726$\pm$0.014&-0.062$\pm$0.006&0.115$\pm$0.018&23.58\\
   2015/07/31&234.7165 &0.6006&4&0.805$\pm$0.012&-0.092$\pm$0.007&-0.072$\pm$0.015&...\\
   2015/08/01&235.7171 &0.6412&4&0.690$\pm$0.011&-0.112$\pm$0.008&-0.126$\pm$0.012&...\\
   2015/08/02&236.7062 &0.6813&4&1.007$\pm$0.021&-0.112$\pm$0.006&-0.031$\pm$0.015&...\\
   2015/08/03&237.7129 &0.7222&4&1.031$\pm$0.018&-0.101$\pm$0.009&-0.012$\pm$0.014&...\\
   2015/08/05&239.6782 &0.8019&4&1.141$\pm$0.032&-0.108$\pm$0.008&-0.002$\pm$0.011&...\\
   2015/08/06&240.7288 &0.8445&4&0.999$\pm$0.015&-0.109$\pm$0.009&-0.004$\pm$0.014&...\\
   2015/08/07&241.7295 &0.8851&4&0.898$\pm$0.012&-0.114$\pm$0.009&-0.009$\pm$0.014&...\\
   2015/09/07&272.6973 &0.1415&5&0.285$\pm$0.021&-0.113$\pm$0.008&-0.205$\pm$0.019&...\\
   2015/09/08&273.6918 &0.1819&5&0.248$\pm$0.015&-0.111$\pm$0.006&-0.185$\pm$0.014&...\\
   2015/09/10&275.6908 &0.2630&5&0.487$\pm$0.033&-0.105$\pm$0.008&-0.107$\pm$0.013&...\\
   2015/09/12&277.6838 &0.3438&5&1.051$\pm$0.020&-0.100$\pm$0.007&0.036$\pm$0.011&45.86\\
   2015/09/13&278.6800 &0.3842&5&1.040$\pm$0.018&-0.094$\pm$0.008&0.051$\pm$0.012&32.03\\
   2015/09/14&279.6804 &0.4248&5&1.038$\pm$0.020&-0.108$\pm$0.007&0.006$\pm$0.016&271.76\\
   2015/09/15&280.6780 &0.4653&5&1.284$\pm$0.016&-0.093$\pm$0.006&0.054$\pm$0.011&37.35\\
   2015/09/16&281.6706 &0.5056&5&1.576$\pm$0.023&-0.075$\pm$0.005&0.127$\pm$0.015&19.49\\
   2015/09/17&282.6714 &0.5462&5&1.330$\pm$0.014&-0.073$\pm$0.006&0.073$\pm$0.011&28.62\\
\end{tabular}
\end{table*}
\begin{table*}
\centering
\contcaption{}
\tabcolsep 0.25cm
\begin{tabular}{@{}cccccccc}
\hline
\bf{Date} & \bf{HJD} & \bf{Phase} & \bf{Cycle} & \multicolumn{3}{c}{\bf{EW({\AA})}}&$\frac{\bf{E_{H{\alpha}}}}{\bf{E_{H{\beta}}}}$\\
\cline{5-7}\\
 & (2,457,000+) & && H$_{\alpha}$ & $\mbox{He~{\sc i}}$ D$_{3}$ & H$_{\beta}$\\
\hline
   2015/09/20&285.7102 &0.6694&5&1.154$\pm$0.018&-0.086$\pm$0.008&0.073$\pm$0.011&24.83\\
   2015/09/23&288.7111 &0.7912&5&0.803$\pm$0.012&-0.101$\pm$0.005&0.030$\pm$0.013&42.05\\
\hline                          
\end{tabular}
\end{table*}
\begin{table*}
\centering
\caption{Observing log and measurements for the subtraction of $\sigma$~Gem.}
\tabcolsep 0.25cm
\label{tab2}
\begin{tabular}{@{}lcccccr}
\hline
\bf{Date} & \bf{HJD} & \bf{Phase} & \bf{Cycle} & \multicolumn{3}{c}{\bf{EW({\AA})}} \\
\cline{5-7}\\
& (2,457,000+) & & & H$_{\alpha}$ & $\mbox{He~{\sc i}}$ D$_{3}$ & H$_{\beta}$\\
\hline
2015/11/04& 330.5781 &0.1136&1&0.122$\pm$0.007&-0.076$\pm$0.002&-0.160$\pm$0.013\\ 
2015/11/04& 330.5803 &0.1137&1&0.114$\pm$0.008&-0.073$\pm$0.002&-0.145$\pm$0.010\\ 
2015/11/05& 331.6532 &0.1684&1&0.079$\pm$0.007&-0.072$\pm$0.005&-0.140$\pm$0.016\\ 
2015/11/05& 331.6553 &0.1686&1&0.071$\pm$0.004&-0.072$\pm$0.002&-0.140$\pm$0.009\\ 
2015/11/06& 332.6210 &0.2178&1&0.259$\pm$0.010&-0.068$\pm$0.004&-0.095$\pm$0.010\\ 
2015/11/06& 332.6231 &0.2179&1&0.253$\pm$0.009&-0.062$\pm$0.006&-0.085$\pm$0.008\\ 
2015/11/07& 333.6156 &0.2685&1&0.206$\pm$0.004&-0.061$\pm$0.007&-0.086$\pm$0.012\\ 
2015/11/07& 333.6178 &0.2687&1&0.217$\pm$0.004&-0.063$\pm$0.005&-0.087$\pm$0.014\\ 
2015/11/12& 338.6076 &0.5232&1&0.086$\pm$0.013&-0.079$\pm$0.003&-0.164$\pm$0.001\\ 
2015/11/12& 338.6097 &0.5233&1&0.111$\pm$0.015&-0.077$\pm$0.004&-0.156$\pm$0.009\\ 
2015/11/14& 340.5911 &0.6244&1&0.038$\pm$0.023&-0.079$\pm$0.004&-0.163$\pm$0.012\\ 
2015/11/14& 340.5932 &0.6245&1&0.034$\pm$0.017&-0.079$\pm$0.006&-0.169$\pm$0.016\\ 
2015/11/18& 344.5197 &0.8248&1&0.168$\pm$0.018&-0.064$\pm$0.005&-0.105$\pm$0.008\\ 
2015/11/18& 344.5219 &0.8249&1&0.173$\pm$0.017&-0.065$\pm$0.006&-0.101$\pm$0.014\\ 
2015/11/20& 346.6361 &0.9327&1&0.251$\pm$0.010&-0.071$\pm$0.006&-0.109$\pm$0.012\\ 
2015/11/20& 346.6382 &0.9328&1&0.238$\pm$0.015&-0.068$\pm$0.009&-0.129$\pm$0.018\\ 
2015/11/21& 347.6432 &0.9841&1&0.148$\pm$0.004&-0.069$\pm$0.004&-0.153$\pm$0.010\\ 
2015/11/21& 347.6454 &0.9842&1&0.147$\pm$0.004&-0.067$\pm$0.002&-0.158$\pm$0.006\\ 
2015/11/22& 348.6418 &0.0350&2&0.095$\pm$0.001&-0.070$\pm$0.002&-0.161$\pm$0.010\\ 
2015/11/22& 348.6440 &0.0351&2&0.098$\pm$0.001&-0.075$\pm$0.007&-0.161$\pm$0.006\\ 
2015/11/24& 350.6372 &0.1368&2&-0.014$\pm$0.002&-0.068$\pm$0.005&-0.212$\pm$0.013\\ 
2015/11/24& 350.6393 &0.1369&2&-0.010$\pm$0.001&-0.067$\pm$0.010&-0.210$\pm$0.009\\ 
2015/11/25& 351.6436 &0.1881&2&0.143$\pm$0.012&-0.058$\pm$0.006&-0.132$\pm$0.012\\ 
2015/11/25& 351.6457 &0.1882&2&0.140$\pm$0.001&-0.058$\pm$0.006&-0.135$\pm$0.014\\ 
2015/11/26& 352.6456 &0.2392&2&0.187$\pm$0.016&-0.059$\pm$0.011&-0.110$\pm$0.007\\ 
2015/11/26& 352.6478 &0.2394&2&0.192$\pm$0.017&-0.059$\pm$0.006&-0.116$\pm$0.013\\ 
2015/11/27& 353.5706 &0.2864&2&0.213$\pm$0.002&-0.057$\pm$0.005&-0.105$\pm$0.011\\ 
2015/11/27& 353.5728 &0.2865&2&0.202$\pm$0.007&-0.060$\pm$0.004&-0.108$\pm$0.015\\ 
2015/11/28& 354.7725 &0.3477&2&0.367$\pm$0.003&-0.068$\pm$0.004&-0.356$\pm$0.008\\ 
2015/11/28& 354.7746 &0.3478&2&0.357$\pm$0.001&-0.068$\pm$0.006&-0.360$\pm$0.012\\ 
2015/11/30& 356.7776 &0.4500&2&0.171$\pm$0.024&-0.087$\pm$0.005&-0.116$\pm$0.012\\ 
2015/11/30& 356.7798 &0.4501&2&0.165$\pm$0.022&-0.092$\pm$0.009&-0.122$\pm$0.009\\ 
2015/12/01& 357.7851 &0.5014&2&0.377$\pm$0.002&-0.080$\pm$0.006&-0.105$\pm$0.014\\ 
2015/12/01& 357.7872 &0.5015&2&0.389$\pm$0.010&-0.079$\pm$0.006&-0.109$\pm$0.010\\ 
2015/12/02& 358.7807 &0.5522&2&0.324$\pm$0.001&-0.081$\pm$0.008&-0.138$\pm$0.006\\ 
2015/12/02& 358.7829 &0.5523&2&0.309$\pm$0.002&-0.082$\pm$0.004&-0.132$\pm$0.010\\ 
2015/12/03& 359.7814 &0.6032&2&0.263$\pm$0.001&-0.074$\pm$0.007&-0.132$\pm$0.008\\ 
2015/12/03& 359.7835 &0.6033&2&0.273$\pm$0.002&-0.076$\pm$0.002&-0.138$\pm$0.012\\ 
2015/12/05& 361.7947 &0.7059&2&0.136$\pm$0.010&-0.091$\pm$0.012&-0.119$\pm$0.010\\ 
2015/12/10& 366.7795 &0.9602&2&0.217$\pm$0.015&-0.066$\pm$0.004&-0.138$\pm$0.017\\ 
2015/12/10& 366.7817 &0.9603&2&0.225$\pm$0.005&-0.064$\pm$0.003&-0.127$\pm$0.011\\ 
2015/12/11& 367.8066 &0.0126&3&0.240$\pm$0.003&-0.084$\pm$0.003&-0.129$\pm$0.012\\ 
2015/12/11& 367.8088 &0.0127&3&0.239$\pm$0.002&-0.082$\pm$0.006&-0.140$\pm$0.009\\ 
2015/12/12& 368.7892 &0.0627&3&0.110$\pm$0.010&-0.092$\pm$0.009&-0.156$\pm$0.012\\ 
2015/12/12& 368.7913 &0.0628&3&0.125$\pm$0.009&-0.091$\pm$0.007&-0.158$\pm$0.012\\ 
2015/12/13& 369.8131 &0.1149&3&0.111$\pm$0.003&-0.082$\pm$0.005&-0.144$\pm$0.016\\ 
2015/12/13& 369.8153 &0.1150&3&0.094$\pm$0.003&-0.088$\pm$0.004&-0.136$\pm$0.009\\ 
2015/12/14& 370.7568 &0.1631&3&0.161$\pm$0.008&-0.078$\pm$0.005&-0.132$\pm$0.008\\ 
2015/12/14& 370.7589 &0.1632&3&0.158$\pm$0.008&-0.084$\pm$0.012&-0.131$\pm$0.010\\ 
2015/12/15& 371.7704 &0.2148&3&0.128$\pm$0.008&-0.067$\pm$0.004&-0.123$\pm$0.008\\ 
2015/12/15& 371.7726 &0.2149&3&0.136$\pm$0.005&-0.068$\pm$0.004&-0.120$\pm$0.012\\ 
2015/12/17& 373.7818 &0.3174&3&0.106$\pm$0.018&-0.070$\pm$0.003&-0.136$\pm$0.010\\ 
2015/12/17& 373.7840 &0.3175&3&0.088$\pm$0.004&-0.069$\pm$0.005&-0.132$\pm$0.014\\ 
2015/12/18& 374.7475 &0.3666&3&0.148$\pm$0.009&-0.060$\pm$0.007&-0.110$\pm$0.012\\ 
2015/12/18& 374.7497 &0.3667&3&0.150$\pm$0.007&-0.065$\pm$0.005&-0.109$\pm$0.009\\ 
2015/12/19& 375.7445 &0.4175&3&0.085$\pm$0.004&-0.068$\pm$0.006&-0.132$\pm$0.011\\ 
2015/12/19& 375.7466 &0.4176&3&0.081$\pm$0.001&-0.074$\pm$0.010&-0.130$\pm$0.011\\ 
2015/12/20& 376.6624 &0.4643&3&0.164$\pm$0.021&-0.058$\pm$0.004&-0.122$\pm$0.012\\ 
2015/12/20& 376.6646 &0.4644&3&0.165$\pm$0.019&-0.058$\pm$0.006&-0.129$\pm$0.010\\ 
2015/12/22& 378.7365 &0.5701&3&0.104$\pm$0.008&-0.071$\pm$0.004&-0.149$\pm$0.009\\ 
2015/12/22& 378.7387 &0.5702&3&0.111$\pm$0.020&-0.073$\pm$0.005&-0.152$\pm$0.011\\ 
2015/12/23& 379.7755 &0.6231&3&0.058$\pm$0.002&-0.067$\pm$0.011&-0.122$\pm$0.014\\ 
\end{tabular}
\end{table*}
\begin{table*}
\centering
\contcaption{}
\tabcolsep 0.25cm
\begin{tabular}{@{}lcccccr}
\hline
\bf{Date} & \bf{HJD} & \bf{Phase} & \bf{Cycle} & \multicolumn{3}{c}{\bf{EW({\AA})}} \\
\cline{5-7}\\
& (2,457,000+) & & & H$_{\alpha}$ & $\mbox{He~{\sc i}}$ D$_{3}$ & H$_{\beta}$\\
\hline
2015/12/23& 379.7776 &0.6232&3&0.061$\pm$0.003&-0.069$\pm$0.002&-0.129$\pm$0.017\\ 
2015/12/24& 380.7284 &0.6717&3&0.064$\pm$0.004&-0.067$\pm$0.002&-0.149$\pm$0.015\\ 
2015/12/24& 380.7306 &0.6718&3&0.063$\pm$0.005&-0.066$\pm$0.005&-0.152$\pm$0.015\\ 
2015/12/28& 384.7176 &0.8752&3&0.271$\pm$0.017&-0.088$\pm$0.004&-0.122$\pm$0.014\\ 
2015/12/28& 384.7197 &0.8753&3&0.281$\pm$0.023&-0.097$\pm$0.009&-0.117$\pm$0.012\\ 
2015/12/29& 385.7153 &0.9261&3&0.203$\pm$0.003&-0.093$\pm$0.007&-0.166$\pm$0.013\\ 
2015/12/29& 385.7174 &0.9262&3&0.202$\pm$0.003&-0.086$\pm$0.010&-0.169$\pm$0.015\\ 
2015/12/30& 386.7125 &0.9769&3&0.140$\pm$0.001&-0.090$\pm$0.008&-0.170$\pm$0.011\\ 
2015/12/30& 386.7143 &0.9770&3&0.134$\pm$0.001&-0.084$\pm$0.014&-0.169$\pm$0.017\\ 
2015/12/31& 387.7104 &0.0279&4&0.185$\pm$0.007&-0.080$\pm$0.010&-0.159$\pm$0.009\\ 
2015/12/31& 387.7125 &0.0280&4&0.178$\pm$0.011&-0.088$\pm$0.008&-0.159$\pm$0.011\\ 
2016/01/04& 391.6334 &0.2280&4&0.286$\pm$0.012&-0.066$\pm$0.011&-0.095$\pm$0.010\\ 
2016/01/04& 391.6355 &0.2281&4&0.284$\pm$0.010&-0.066$\pm$0.008&-0.094$\pm$0.012\\
2016/01/05& 392.6481 &0.2797&4&0.267$\pm$0.003&-0.054$\pm$0.012&-0.086$\pm$0.015\\
2016/01/06& 393.6253 &0.3296&4&0.554$\pm$0.010&-0.037$\pm$0.008&0.147$\pm$0.025\\
2016/01/06& 393.6275 &0.3297&4&0.587$\pm$0.006&-0.039$\pm$0.006&0.150$\pm$0.026\\
2016/01/06& 394.3565 &0.3669&4&0.410$\pm$0.003&-0.058$\pm$0.005&-0.023$\pm$0.012\\
2016/01/06& 394.3587 &0.3670&4&0.412$\pm$0.022&-0.059$\pm$0.002&-0.025$\pm$0.010\\
2016/01/07& 395.3620 &0.4182&4&0.297$\pm$0.019&-0.063$\pm$0.002&-0.061$\pm$0.015\\
2016/01/07& 395.3641 &0.4183&4&0.292$\pm$0.018&-0.049$\pm$0.004&-0.056$\pm$0.011\\
2016/01/08& 396.4219 &0.4722&4&0.254$\pm$0.028&-0.073$\pm$0.006&-0.089$\pm$0.010\\
2016/01/08& 396.4241 &0.4723&4&0.256$\pm$0.030&-0.072$\pm$0.004&-0.091$\pm$0.014\\ 
2016/01/09& 397.4292 &0.5236&4&0.119$\pm$0.010&-0.089$\pm$0.005&-0.144$\pm$0.009\\ 
2016/01/09& 397.4314 &0.5237&4&0.110$\pm$0.011&-0.085$\pm$0.003&-0.139$\pm$0.015\\ 
2016/01/11& 399.3877 &0.6235&4&0.083$\pm$0.001&-0.089$\pm$0.002&-0.181$\pm$0.014\\ 
2016/01/11& 399.3898 &0.6236&4&0.091$\pm$0.002&-0.088$\pm$0.002&-0.185$\pm$0.012\\ 
2016/01/12& 400.4178 &0.6760&4&-0.054$\pm$0.005&-0.085$\pm$0.004&-0.230$\pm$0.010\\ 
2016/01/12& 400.4200 &0.6761&4&-0.066$\pm$0.001&-0.091$\pm$0.006&-0.228$\pm$0.011\\ 
2016/01/12& 400.4221 &0.6763&4&-0.060$\pm$0.002&-0.086$\pm$0.003&-0.231$\pm$0.009\\ 
2016/01/12& 400.4243 &0.6764&4&-0.062$\pm$0.001&-0.086$\pm$0.002&-0.236$\pm$0.011\\ 
2016/01/12& 400.4264 &0.6765&4&-0.050$\pm$0.002&-0.081$\pm$0.004&-0.236$\pm$0.010\\ 
2016/01/12& 400.4292 &0.6766&4&-0.058$\pm$0.001&-0.082$\pm$0.002&-0.234$\pm$0.012\\ 
2016/01/12& 400.4314 &0.6767&4&-0.064$\pm$0.004&-0.078$\pm$0.002&-0.233$\pm$0.008\\ 
2016/01/12& 400.4335 &0.6768&4&-0.059$\pm$0.003&-0.080$\pm$0.003&-0.232$\pm$0.006\\ 
2016/01/12& 400.4357 &0.6770&4&-0.068$\pm$0.004&-0.084$\pm$0.004&-0.231$\pm$0.008\\ 
2016/01/12& 400.4378 &0.6771&4&-0.058$\pm$0.001&-0.080$\pm$0.004&-0.239$\pm$0.010\\ 
2016/01/13& 401.3822 &0.7252&4&0.068$\pm$0.002&-0.091$\pm$0.009&-0.149$\pm$0.006\\ 
2016/01/13& 401.3843 &0.7253&4&0.065$\pm$0.003&-0.084$\pm$0.006&-0.153$\pm$0.012\\ 
2016/01/13& 401.3865 &0.7255&4&0.065$\pm$0.006&-0.082$\pm$0.004&-0.159$\pm$0.014\\ 
2016/01/13& 401.3886 &0.7256&4&0.044$\pm$0.001&-0.084$\pm$0.002&-0.152$\pm$0.017\\ 
2016/01/13& 401.3908 &0.7257&4&0.058$\pm$0.003&-0.080$\pm$0.004&-0.159$\pm$0.017\\ 
2016/01/13& 401.4480 &0.7286&4&0.072$\pm$0.009&-0.077$\pm$0.005&-0.156$\pm$0.014\\ 
2016/01/13& 401.4501 &0.7287&4&0.070$\pm$0.006&-0.074$\pm$0.004&-0.158$\pm$0.008\\ 
2016/01/13& 401.4523 &0.7288&4&0.095$\pm$0.005&...&...\\ 
2016/01/13& 401.4544 &0.7289&4&0.072$\pm$0.003&...&...\\ 
2016/01/13& 401.4566 &0.7290&4&0.067$\pm$0.013&...&...\\ 
2016/01/14& 402.4248 &0.7784&4&0.184$\pm$0.001&-0.082$\pm$0.006&-0.105$\pm$0.015\\ 
2016/01/14& 402.4270 &0.7785&4&0.191$\pm$0.005&-0.092$\pm$0.008&-0.129$\pm$0.005\\ 
2016/01/14& 402.4291 &0.7786&4&0.178$\pm$0.005&-0.088$\pm$0.004&-0.130$\pm$0.015\\ 
2016/01/14& 402.4313 &0.7787&4&0.186$\pm$0.006&-0.088$\pm$0.004&-0.136$\pm$0.010\\ 
2016/01/14& 402.4334 &0.7789&4&0.193$\pm$0.007&-0.083$\pm$0.004&-0.127$\pm$0.011\\ 
2016/01/14& 402.4361 &0.7790&4&0.166$\pm$0.006&-0.085$\pm$0.002&-0.135$\pm$0.015\\ 
2016/01/15& 403.4297 &0.8297&4&0.131$\pm$0.001&-0.091$\pm$0.007&-0.171$\pm$0.020\\ 
2016/01/15& 403.4361 &0.8300&4&0.146$\pm$0.020&...&...\\ 
2016/01/15& 403.4389 &0.8301&4&0.133$\pm$0.021&...&...\\ 
2016/01/17& 404.5074 &0.8847&4&0.215$\pm$0.019&-0.083$\pm$0.008&-0.146$\pm$0.012\\ 
2016/01/17& 404.5096 &0.8848&4&0.214$\pm$0.019&-0.079$\pm$0.005&-0.144$\pm$0.012\\ 
2016/01/17& 404.5117 &0.8849&4&0.226$\pm$0.027&-0.078$\pm$0.006&-0.145$\pm$0.010\\ 
2016/01/17& 404.5139 &0.8850&4&0.223$\pm$0.023&-0.077$\pm$0.004&-0.148$\pm$0.011\\ 
2016/01/17& 404.5160 &0.8851&4&0.219$\pm$0.022&-0.078$\pm$0.005&-0.141$\pm$0.006\\ 
2016/01/17& 404.5191 &0.8852&4&0.229$\pm$0.034&-0.071$\pm$0.004&-0.145$\pm$0.006\\ 
2016/01/17& 404.5213 &0.8854&4&0.229$\pm$0.025&-0.075$\pm$0.005&-0.148$\pm$0.005\\ 
\end{tabular}
\end{table*}
\begin{table*}
\centering
\contcaption{}
\tabcolsep 0.25cm
\begin{tabular}{@{}lcccccr}
\hline
\bf{Date} & \bf{HJD} & \bf{Phase} & \bf{Cycle} & \multicolumn{3}{c}{\bf{EW({\AA})}} \\
\cline{5-7}\\
& (2,457,000+) & & & H$_{\alpha}$ & $\mbox{He~{\sc i}}$ D$_{3}$ & H$_{\beta}$\\
\hline
2016/01/17& 404.5234 &0.8855&4&0.229$\pm$0.022&-0.073$\pm$0.007&-0.144$\pm$0.008\\ 
2016/01/17& 404.5256 &0.8856&4&0.226$\pm$0.024&-0.069$\pm$0.008&-0.146$\pm$0.004\\ 
2016/01/17& 404.5277 &0.8857&4&0.225$\pm$0.021&-0.072$\pm$0.006&-0.144$\pm$0.010\\ 
2016/01/18& 406.3401 &0.9781&4&0.127$\pm$0.001&-0.075$\pm$0.006&-0.191$\pm$0.009\\ 
2016/01/18& 406.3422 &0.9782&4&0.120$\pm$0.001&-0.076$\pm$0.004&-0.190$\pm$0.013\\ 
2016/01/18& 406.3444 &0.9784&4&0.124$\pm$0.001&-0.073$\pm$0.006&-0.179$\pm$0.012\\ 
2016/01/18& 406.3465 &0.9785&4&0.120$\pm$0.002&-0.074$\pm$0.002&-0.188$\pm$0.009\\ 
2016/01/18& 406.3487 &0.9786&4&0.128$\pm$0.017&-0.071$\pm$0.004&-0.194$\pm$0.010\\ 
2016/01/18& 406.3515 &0.9787&4&0.126$\pm$0.014&-0.069$\pm$0.005&-0.194$\pm$0.010\\ 
2016/01/18& 406.3537 &0.9788&4&0.127$\pm$0.012&-0.068$\pm$0.005&-0.183$\pm$0.014\\ 
2016/01/18& 406.3558 &0.9789&4&0.134$\pm$0.014&-0.072$\pm$0.004&-0.189$\pm$0.012\\ 
2016/01/18& 406.3580 &0.9790&4&0.128$\pm$0.016&-0.076$\pm$0.008&-0.192$\pm$0.011\\ 
2016/01/19& 407.3729 &0.0308&5&0.145$\pm$0.003&-0.071$\pm$0.006&-0.173$\pm$0.009\\ 
2016/01/19& 407.3750 &0.0309&5&0.140$\pm$0.002&-0.064$\pm$0.012&-0.175$\pm$0.015\\ 
2016/01/19& 407.3772 &0.0310&5&0.146$\pm$0.001&-0.071$\pm$0.005&-0.175$\pm$0.012\\ 
2016/01/19& 407.3793 &0.0311&5&0.145$\pm$0.002&-0.072$\pm$0.002&-0.166$\pm$0.014\\ 
2016/01/19& 407.3815 &0.0313&5&0.144$\pm$0.001&-0.071$\pm$0.004&-0.167$\pm$0.015\\ 
2016/01/20& 408.3569 &0.0810&5&0.122$\pm$0.003&-0.075$\pm$0.004&-0.176$\pm$0.010\\ 
2016/01/20& 408.3591 &0.0811&5&0.122$\pm$0.001&-0.072$\pm$0.006&-0.174$\pm$0.012\\ 
2016/01/20& 408.3612 &0.0812&5&0.119$\pm$0.001&-0.084$\pm$0.003&-0.169$\pm$0.014\\ 
2016/01/20& 408.3634 &0.0813&5&0.125$\pm$0.003&-0.086$\pm$0.004&-0.174$\pm$0.009\\ 
2016/01/20& 408.3656 &0.0814&5&0.125$\pm$0.011&-0.087$\pm$0.005&-0.174$\pm$0.008\\ 
2016/01/20& 408.3683 &0.0816&5&0.122$\pm$0.007&-0.082$\pm$0.005&-0.177$\pm$0.008\\ 
2016/01/20& 408.3705 &0.0817&5&0.126$\pm$0.015&-0.080$\pm$0.004&-0.170$\pm$0.015\\ 
2016/01/20& 408.3726 &0.0818&5&0.124$\pm$0.013&-0.086$\pm$0.006&-0.176$\pm$0.012\\ 
2016/01/20& 408.3748 &0.0819&5&0.118$\pm$0.007&-0.085$\pm$0.001&-0.171$\pm$0.013\\ 
2016/01/20& 408.3769 &0.0820&5&0.115$\pm$0.008&-0.080$\pm$0.003&-0.167$\pm$0.016\\ 
2016/01/21& 409.3804 &0.1332&5&0.137$\pm$0.001&-0.080$\pm$0.003&-0.163$\pm$0.009\\ 
2016/01/21& 409.3825 &0.1333&5&0.131$\pm$0.003&-0.076$\pm$0.006&-0.157$\pm$0.008\\ 
2016/01/21& 409.3847 &0.1334&5&0.130$\pm$0.002&-0.080$\pm$0.002&-0.160$\pm$0.015\\ 
2016/01/23& 411.3222 &0.2323&5&0.201$\pm$0.016&-0.060$\pm$0.004&-0.131$\pm$0.009\\ 
2016/01/23& 411.3244 &0.2324&5&0.214$\pm$0.013&-0.065$\pm$0.002&-0.129$\pm$0.011\\ 
2016/01/23& 411.3265 &0.2325&5&0.225$\pm$0.016&-0.064$\pm$0.002&-0.122$\pm$0.008\\ 
2016/01/23& 411.3287 &0.2326&5&0.225$\pm$0.019&-0.062$\pm$0.005&-0.130$\pm$0.008\\ 
2016/01/23& 411.3308 &0.2327&5&0.217$\pm$0.016&-0.066$\pm$0.003&-0.125$\pm$0.015\\ 
2016/01/23& 411.3335 &0.2328&5&0.221$\pm$0.014&-0.069$\pm$0.004&-0.132$\pm$0.012\\ 
2016/01/23& 411.3356 &0.2329&5&0.212$\pm$0.012&-0.059$\pm$0.005&-0.131$\pm$0.010\\ 
2016/01/23& 411.3378 &0.2331&5&0.227$\pm$0.013&-0.068$\pm$0.004&-0.121$\pm$0.012\\ 
2016/01/23& 411.3399 &0.2332&5&0.207$\pm$0.012&-0.060$\pm$0.006&-0.127$\pm$0.008\\ 
2016/01/23& 411.3421 &0.2333&5&0.214$\pm$0.010&-0.064$\pm$0.004&-0.126$\pm$0.009\\ 
2016/01/24& 412.3378 &0.2841&5&0.181$\pm$0.004&...&...\\ 
2016/01/24& 412.3405 &0.2842&5&0.172$\pm$0.017&-0.061$\pm$0.004&...\\ 
2016/01/24& 412.3427 &0.2843&5&0.153$\pm$0.001&-0.071$\pm$0.005&...\\ 
2016/01/25& 413.3502 &0.3357&5&0.177$\pm$0.002&-0.060$\pm$0.004&-0.118$\pm$0.017\\ 
2016/01/25& 413.3522 &0.3358&5&0.178$\pm$0.001&-0.062$\pm$0.002&-0.129$\pm$0.015\\ 
2016/01/25& 413.3544 &0.3359&5&0.184$\pm$0.001&-0.060$\pm$0.001&-0.130$\pm$0.015\\ 
2016/01/25& 413.3565 &0.3360&5&0.178$\pm$0.001&-0.062$\pm$0.002&-0.133$\pm$0.016\\ 
2016/01/26& 414.3517 &0.3868&5&0.133$\pm$0.002&-0.073$\pm$0.004&-0.119$\pm$0.012\\ 
2016/01/26& 414.3538 &0.3869&5&0.139$\pm$0.002&-0.077$\pm$0.005&-0.120$\pm$0.009\\ 
2016/01/26& 414.3560 &0.3870&5&0.140$\pm$0.001&-0.072$\pm$0.009&-0.128$\pm$0.011\\ 
2016/01/26& 414.3581 &0.3871&5&0.127$\pm$0.002&-0.069$\pm$0.006&-0.126$\pm$0.011\\ 
2016/01/26& 414.3603 &0.3872&5&0.139$\pm$0.002&-0.072$\pm$0.004&-0.118$\pm$0.012\\ 
2016/01/26& 414.3630 &0.3874&5&0.137$\pm$0.001&-0.074$\pm$0.005&-0.119$\pm$0.010\\ 
2016/01/26& 414.3652 &0.3875&5&0.141$\pm$0.001&-0.071$\pm$0.003&-0.121$\pm$0.013\\ 
2016/01/26& 414.3673 &0.3876&5&0.139$\pm$0.002&-0.067$\pm$0.006&-0.125$\pm$0.009\\ 
2016/01/26& 414.3695 &0.3877&5&0.136$\pm$0.001&-0.077$\pm$0.006&-0.125$\pm$0.009\\ 
2016/01/26& 414.3716 &0.3878&5&0.138$\pm$0.004&-0.076$\pm$0.004&-0.132$\pm$0.019\\ 
2016/01/27& 415.3517 &0.4378&5&0.115$\pm$0.003&-0.080$\pm$0.002&-0.144$\pm$0.010\\ 
2016/01/27& 415.3539 &0.4379&5&0.113$\pm$0.003&-0.084$\pm$0.002&-0.141$\pm$0.012\\ 
2016/01/27& 415.3560 &0.4380&5&0.113$\pm$0.002&-0.082$\pm$0.003&-0.146$\pm$0.008\\ 
2016/01/27& 415.3582 &0.4381&5&0.109$\pm$0.002&-0.080$\pm$0.005&-0.142$\pm$0.010\\ 
2016/01/27& 415.3603 &0.4382&5&0.114$\pm$0.003&-0.085$\pm$0.003&-0.149$\pm$0.011\\ 
\end{tabular}
\end{table*}
\begin{table*}
\centering
\contcaption{}
\tabcolsep 0.25cm
\begin{tabular}{@{}lcccccr}
\hline
\bf{Date} & \bf{HJD} & \bf{Phase} & \bf{Cycle} & \multicolumn{3}{c}{\bf{EW({\AA})}} \\
\cline{5-7}\\
& (2,457,000+) & & & H$_{\alpha}$ & $\mbox{He~{\sc i}}$ D$_{3}$ & H$_{\beta}$\\
\hline
2016/01/27& 415.3630 &0.4384&5&0.108$\pm$0.005&-0.077$\pm$0.010&-0.144$\pm$0.009\\ 
2016/01/27& 415.3652 &0.4385&5&0.115$\pm$0.002&-0.082$\pm$0.003&-0.150$\pm$0.010\\ 
2016/01/27& 415.3673 &0.4386&5&0.114$\pm$0.003&-0.081$\pm$0.006&-0.145$\pm$0.012\\ 
2016/01/27& 415.3695 &0.4387&5&0.112$\pm$0.001&-0.082$\pm$0.004&-0.145$\pm$0.008\\ 
2016/01/27& 415.3716 &0.4388&5&0.117$\pm$0.002&-0.080$\pm$0.006&-0.148$\pm$0.006\\ 
2016/01/30& 418.3911 &0.5928&5&-0.012$\pm$0.001&-0.074$\pm$0.005&-0.154$\pm$0.030\\ 
2016/02/01& 420.3557 &0.6931&5&0.119$\pm$0.005&-0.079$\pm$0.003&-0.116$\pm$0.015\\ 
2016/02/03& 422.3404 &0.7943&5&0.196$\pm$0.011&-0.092$\pm$0.012&-0.141$\pm$0.018\\ 
2016/02/03& 422.3425 &0.7944&5&0.197$\pm$0.009&-0.090$\pm$0.009&-0.150$\pm$0.010\\ 
2016/02/04& 423.3528 &0.8459&5&0.274$\pm$0.026&-0.094$\pm$0.004&-0.146$\pm$0.012\\ 
2016/02/04& 423.3550 &0.8460&5&0.261$\pm$0.011&-0.092$\pm$0.015&-0.158$\pm$0.008\\ 
2016/02/04& 423.3571 &0.8462&5&0.266$\pm$0.011&-0.092$\pm$0.013&-0.150$\pm$0.010\\ 
2016/02/05& 424.3416 &0.8964&5&0.251$\pm$0.003&-0.090$\pm$0.006&-0.147$\pm$0.012\\ 
2016/02/05& 424.3438 &0.8965&5&0.242$\pm$0.010&-0.082$\pm$0.010&-0.148$\pm$0.015\\ 
2016/02/05& 424.3459 &0.8966&5&0.245$\pm$0.010&-0.072$\pm$0.008&-0.145$\pm$0.010\\ 
2016/02/05& 424.3481 &0.8967&5&0.241$\pm$0.011&-0.077$\pm$0.004&-0.159$\pm$0.012\\ 
2016/02/05& 424.3503 &0.8968&5&0.245$\pm$0.009&-0.076$\pm$0.004&-0.155$\pm$0.010\\ 
2016/02/07& 426.3434 &0.9985&5&0.301$\pm$0.004&-0.076$\pm$0.002&-0.106$\pm$0.019\\ 
2016/02/07& 426.3455 &0.9986&5&0.299$\pm$0.003&-0.072$\pm$0.007&-0.094$\pm$0.017\\ 
2016/02/07& 426.3477 &0.9987&5&0.305$\pm$0.004&-0.068$\pm$0.012&-0.089$\pm$0.012\\ 
2016/02/07& 426.3498 &0.9988&5&0.307$\pm$0.004&-0.074$\pm$0.006&-0.089$\pm$0.011\\ 
2016/02/07& 426.3520 &0.9989&5&0.301$\pm$0.004&-0.077$\pm$0.006&-0.081$\pm$0.015\\ 
2016/02/07& 426.3549 &0.9991&5&0.304$\pm$0.003&-0.076$\pm$0.004&-0.093$\pm$0.017\\ 
2016/02/07& 426.3571 &0.9992&5&0.304$\pm$0.003&-0.079$\pm$0.008&-0.073$\pm$0.019\\ 
2016/02/07& 426.3592 &0.9993&5&0.295$\pm$0.006&-0.061$\pm$0.015&-0.094$\pm$0.015\\ 
2016/02/07& 426.3614 &0.9994&5&0.303$\pm$0.004&-0.073$\pm$0.005&-0.089$\pm$0.010\\ 
2016/02/07& 426.3635 &0.9995&5&0.294$\pm$0.004&-0.074$\pm$0.008&-0.091$\pm$0.018\\ 
2016/02/09& 428.3431 &0.1005&6&0.263$\pm$0.002&-0.075$\pm$0.006&-0.117$\pm$0.012\\ 
2016/02/09& 428.3453 &0.1006&6&0.267$\pm$0.006&-0.069$\pm$0.009&-0.118$\pm$0.011\\ 
2016/02/09& 428.3474 &0.1007&6&0.265$\pm$0.005&-0.080$\pm$0.011&-0.118$\pm$0.011\\ 
2016/02/09& 428.3496 &0.1008&6&0.267$\pm$0.004&-0.079$\pm$0.006&-0.115$\pm$0.013\\ 
2016/02/09& 428.3517 &0.1009&6&0.269$\pm$0.005&-0.074$\pm$0.008&-0.121$\pm$0.005\\ 
2016/02/09& 428.3547 &0.1011&6&0.266$\pm$0.006&-0.079$\pm$0.004&-0.121$\pm$0.008\\ 
2016/02/09& 428.3568 &0.1012&6&0.265$\pm$0.004&-0.079$\pm$0.002&-0.125$\pm$0.004\\ 
2016/02/09& 428.3590 &0.1013&6&0.264$\pm$0.004&-0.078$\pm$0.002&-0.124$\pm$0.006\\ 
2016/02/09& 428.3611 &0.1014&6&0.270$\pm$0.005&-0.079$\pm$0.004&-0.127$\pm$0.006\\ 
2016/02/09& 428.3633 &0.1015&6&0.274$\pm$0.003&-0.076$\pm$0.003&-0.122$\pm$0.008\\ 
2016/02/11& 430.3365 &0.2022&6&0.039$\pm$0.007&-0.064$\pm$0.006&-0.224$\pm$0.010\\ 
2016/02/11& 430.3386 &0.2023&6&0.042$\pm$0.006&-0.067$\pm$0.003&-0.225$\pm$0.015\\ 
2016/02/11& 430.3408 &0.2024&6&0.032$\pm$0.005&-0.067$\pm$0.004&-0.230$\pm$0.012\\ 
2016/02/11& 430.3429 &0.2025&6&0.043$\pm$0.004&-0.066$\pm$0.002&-0.221$\pm$0.009\\ 
2016/02/11& 430.3451 &0.2026&6&0.043$\pm$0.008&-0.067$\pm$0.002&-0.224$\pm$0.010\\ 
2016/02/11& 430.3477 &0.2027&6&0.037$\pm$0.004&-0.068$\pm$0.004&-0.226$\pm$0.006\\ 
2016/02/11& 430.3499 &0.2028&6&0.043$\pm$0.004&-0.067$\pm$0.003&-0.227$\pm$0.006\\ 
2016/02/11& 430.3520 &0.2030&6&0.038$\pm$0.004&-0.069$\pm$0.005&-0.228$\pm$0.006\\ 
2016/02/11& 430.3542 &0.2031&6&0.033$\pm$0.005&-0.063$\pm$0.006&-0.224$\pm$0.009\\ 
2016/02/11& 430.3563 &0.2032&6&0.047$\pm$0.002&-0.070$\pm$0.009&-0.230$\pm$0.011\\ 
2016/02/12& 431.3261 &0.2526&6&0.091$\pm$0.004&-0.074$\pm$0.008&-0.165$\pm$0.013\\ 
2016/02/12& 431.3282 &0.2527&6&0.098$\pm$0.002&-0.076$\pm$0.006&-0.155$\pm$0.008\\ 
2016/02/14& 433.3368 &0.3552&6&0.013$\pm$0.001&-0.067$\pm$0.009&-0.179$\pm$0.010\\ 
2016/02/14& 433.3389 &0.3553&6&0.010$\pm$0.001&-0.071$\pm$0.006&-0.169$\pm$0.012\\ 
2016/02/14& 433.3411 &0.3554&6&0.014$\pm$0.001&-0.072$\pm$0.006&-0.167$\pm$0.012\\ 
2016/02/14& 433.3432 &0.3555&6&0.010$\pm$0.001&-0.069$\pm$0.013&-0.179$\pm$0.014\\ 
2016/02/16& 435.3203 &0.4564&6&0.066$\pm$0.001&-0.053$\pm$0.008&-0.149$\pm$0.009\\ 
2016/02/17& 436.3227 &0.5075&6&0.112$\pm$0.004&-0.057$\pm$0.009&-0.143$\pm$0.011\\ 
2016/02/17& 436.3249 &0.5076&6&0.118$\pm$0.006&-0.066$\pm$0.010&-0.149$\pm$0.008\\
2016/02/17& 436.3270 &0.5077&6&0.115$\pm$0.005&-0.064$\pm$0.004&-0.154$\pm$0.005\\ 
2016/02/17& 436.3292 &0.5078&6&0.115$\pm$0.006&-0.063$\pm$0.006&-0.152$\pm$0.004\\ 
2016/02/26& 444.6418 &0.9319&6&0.295$\pm$0.005&-0.088$\pm$0.009&-0.115$\pm$0.008\\ 
2016/02/26& 444.6440 &0.9320&6&0.295$\pm$0.004&-0.077$\pm$0.011&-0.119$\pm$0.004\\ 
2016/02/26& 444.6461 &0.9321&6&0.301$\pm$0.003&-0.083$\pm$0.006&-0.115$\pm$0.011\\  
2016/03/01& 448.6459 &0.1361&7&0.084$\pm$0.011&-0.075$\pm$0.003&-0.205$\pm$0.010\\ 
\end{tabular}
\end{table*}
\begin{table*}
\centering
\contcaption{}
\tabcolsep 0.25cm
\begin{tabular}{@{}lcccccr}
\hline
\bf{Date} & \bf{HJD} & \bf{Phase} & \bf{Cycle} & \multicolumn{3}{c}{\bf{EW({\AA})}} \\
\cline{5-7}\\
& (2,457,000+) & & & H$_{\alpha}$ & $\mbox{He~{\sc i}}$ D$_{3}$ & H$_{\beta}$\\
\hline
2016/03/04& 451.6303 &0.2883&7&-0.060$\pm$0.002&-0.076$\pm$0.005&-0.203$\pm$0.008\\ 
2016/03/04& 451.6324 &0.2884&7&-0.073$\pm$0.001&-0.078$\pm$0.006&-0.200$\pm$0.014\\ 
2016/03/04& 451.6346 &0.2885&7&-0.063$\pm$0.003&-0.075$\pm$0.004&-0.202$\pm$0.009\\ 
2016/03/04& 451.6367 &0.2887&7&-0.061$\pm$0.002&-0.071$\pm$0.004&-0.205$\pm$0.011\\ 
2016/03/06& 454.3367 &0.4264&7&-0.052$\pm$0.005&-0.049$\pm$0.017&-0.219$\pm$0.007\\ 
2016/03/06& 454.3389 &0.4265&7&-0.046$\pm$0.005&-0.060$\pm$0.005&-0.224$\pm$0.008\\ 
2016/03/06& 454.3410 &0.4266&7&-0.052$\pm$0.006&-0.060$\pm$0.004&-0.221$\pm$0.010\\ 
2016/03/10& 458.3223 &0.6297&7&0.346$\pm$0.011&-0.067$\pm$0.002&-0.102$\pm$0.007\\ 
2016/03/10& 458.3244 &0.6298&7&0.328$\pm$0.003&-0.061$\pm$0.002&-0.095$\pm$0.009\\ 
2016/03/11& 459.3589 &0.6826&7&0.075$\pm$0.001&-0.098$\pm$0.007&-0.171$\pm$0.006\\ 
2016/03/11& 459.3610 &0.6827&7&0.074$\pm$0.002&-0.108$\pm$0.011&-0.160$\pm$0.015\\ 
2016/03/13& 461.3232 &0.7828&7&0.095$\pm$0.001&-0.087$\pm$0.009&-0.178$\pm$0.008\\ 
2016/03/13& 461.3253 &0.7829&7&0.084$\pm$0.002&-0.096$\pm$0.008&-0.178$\pm$0.008\\ 
2016/03/17& 465.3244 &0.9868&7&0.155$\pm$0.016&-0.064$\pm$0.005&-0.162$\pm$0.010\\ 
2016/03/17& 465.3265 &0.9870&7&0.156$\pm$0.012&-0.068$\pm$0.005&-0.153$\pm$0.017\\ 
2016/03/19& 467.3303 &0.0892&8&0.171$\pm$0.001&-0.084$\pm$0.006&-0.187$\pm$0.011\\ 
2016/03/19& 467.3324 &0.0893&8&0.170$\pm$0.001&-0.083$\pm$0.004&-0.186$\pm$0.009\\ 
2016/03/24& 472.3264 &0.3440&8&0.096$\pm$0.005&-0.065$\pm$0.009&-0.164$\pm$0.010\\ 
2016/03/24& 472.3286 &0.3441&8&0.097$\pm$0.002&-0.060$\pm$0.005&-0.163$\pm$0.008\\ 
2016/03/26& 474.3272 &0.4461&8&0.152$\pm$0.020&-0.044$\pm$0.010&-0.157$\pm$0.012\\ 
2016/03/26& 474.3294 &0.4462&8&0.150$\pm$0.020&-0.049$\pm$0.005&-0.149$\pm$0.011\\ 
2016/04/01& 480.4128 &0.7565&8&0.151$\pm$0.023&-0.086$\pm$0.004&-0.158$\pm$0.009\\ 
2016/04/01& 480.4149 &0.7566&8&0.151$\pm$0.023&-0.089$\pm$0.006&-0.150$\pm$0.013\\ 
\hline                          
\end{tabular}
\end{table*}

\bsp	
\label{lastpage}
\end{document}